\documentclass{article}
\usepackage{threeparttable}
\usepackage{authblk}
\usepackage{graphicx} 
\usepackage[left=1.5in, right=1.5in, top=1.5in, bottom=1.5in]{geometry}
\usepackage[natbibapa]{apacite}
\usepackage{xcolor}
\usepackage{adjustbox}
\usepackage{float}
\usepackage{subcaption}
\usepackage[hidelinks]{hyperref}
\usepackage{booktabs}
\usepackage{amsmath}

\title{Co-evolution of the global research collaboration network and the performance of nations in science and technology}

\author[1]{Travis A. Whetsell}
\author[1,2]{Jeongyoon Jane Yang}

\affil[1]{Jimmy and Rosalynn Carter School of Public Policy, Georgia Institute of Technology, Atlanta, Georgia}
\affil[2]{Andrew Young School of Policy Studies, Georgia State University, Atlanta, Georgia}

\date{\today}

\begin{document}

\maketitle

\begin{abstract}

Researchers have long suspected that international research collaboration (IRC) and scientific and technological (S\&T) performance are subject to reciprocal causality, yet the endogenous co-evolution of these twin phenomena has yet to be tested by large-scale empirical analysis. This study tests these effects simultaneously using a longitudinal co-evolution model on three decades of global network and national performance data. Stochastic actor-oriented models (SAOM) are used to analyze data on 172 countries from 1993 to 2022. Yearly IRC networks are constructed from Web of Science's XML database, and performance data are gathered from Elsevier's fractional field-weighted citation impact (FWCI). The models also account for geographic, economic, demographic, and political factors, as well as endogenous network processes. The results provide support for co-evolution. Distance and shared language moderate this relationship in contrasting ways. The selection effect of performance on tie formation is amplified across distance and dampened by a shared language, whereas the influence effect of performance on centrality is attenuated by remoteness and strengthened by linguistic reach. This pattern suggests that information asymmetry shapes partner selection, while communication and coordination shape the returns to collaboration, pointing to a signaling role for citation-based performance metrics in collaborator selection.

\end{abstract}

\section*{\small \centering Acknowledgments} {\small } 
Earlier versions of this research were presented at the 2025 Atlanta Conference for Science and Innovation Policy and the 2025 International Network for Social Network Analysis Conference. Thank you to Diana Hicks for comments on a draft of the manuscript. Claude Opus 5 and Fable 5, via Claude Code, were used for Python, R, and LaTeX programming, auditing of the mathematical exposition, and general editorial support. The authors take full responsibility for the content.

\newpage

\section{Introduction}

International research collaboration (IRC) has become a defining feature of the production of modern science and technology (S\&T) \citep{wagner2005network, adams2013fourth}. The vision of the heroic lone researcher is no longer a standard feature of S\&T progress. Today, research teams are responsible for producing a large share of scientific progress, and they increasingly span national borders \citep{wuchty2007increasing, leydesdorff2008international, fortunato2018science}. While a large body of literature has emphasized the functional effects of IRC on S\&T research performance, fewer studies have reversed the causal arrow \citep{abramo2009research, abramo2011relationship} and none have examined both pathways simultaneously. Existing research in the emerging IRC research domain has mostly examined single or comparative case studies, relatively small samples covering developed countries, and uni-directional relations between empirical indicators \citep{chen2019international}. Although large scale network studies of IRC are becoming more common, drivers of IRC network formation and effects of the IRC network on policy-relevant performance outputs tend to be analyzed separately. Large scale analysis of the reciprocal dynamic relationship between IRC and national performance has yet to be conducted.  

Despite most of the attention going to the performance side of the coin \citep{glanzel2001national, kwiek2021large, waltman2016review} and a more general lack of research targeting endogenous effects explicitly, a stream of scholarship on social processes, including the Matthew Effect, preferential attachment, and other mechanisms that generate increasing returns to scale, reserve an important place in the body of research on S\&T policy, sociology of science, scientometrics, and informetrics \citep{barabasi2002evolution, merton1968matthew, newman2001structure, price1965networks}. Classic works on the Matthew Effect for example analyze reciprocal effects between social processes and research outputs and outcomes in a mostly conceptual way \citep{merton1968matthew}, and little progress has been made on quantitatively addressing such dynamical processes. Modern computational advancements have dramatically improved the situation \citep{ferligoj2015scientific, fortunato2018science, perc2014matthew}.  

This article takes a tiered approach to hypothesis testing. First, this article advances the core proposition that the global IRC network co-evolves with national S\&T performance over time. Second, it advances two primary hypotheses: 1) that network centrality in the IRC network enhances national S\&T performance over time, 2) national S\&T performance enhances the network centrality of countries in the IRC network. Third, it advances four auxiliary working hypotheses regarding moderation effects of geographic distance and language proximity. The hypotheses are tested simultaneously using longitudinal co-evolution models on three decades of global S\&T network and performance data. The stochastic actor-oriented model (SAOM) framework \citep{snijders2001statistical, snijders2017modeling} is used to analyze data on 172 countries from 1993 to 2022. Yearly IRC networks are constructed from Web of Science's XML database. Corresponding national S\&T performance data is gathered from Elsevier's fractional field-weighted citation impact (FWCI), which disentangles national from internationally attributed citation impact \citep{purkayastha2019comparison, wagner2018openness}. The models also account for geographic and linguistic distance, national wealth, population metrics, political governance, publication volume, and endogenous network processes.   

The results support both primary hypotheses and all four auxiliary hypotheses. The collaboration network and national performance display reciprocal co-evolution. Network centrality appears to enhance performance, while performance appears to enhance country network centrality. Geographic distance and shared language moderate this relationship in contrasting ways. On the selection side, the effect of performance on tie formation is amplified across distance and dampened by a shared language, whereas on the returns side the performance benefit of centrality is attenuated by remoteness and strengthened by linguistic reach. This suggests that information asymmetry shapes partner selection, consistent with a signaling account, in which actors lacking direct information rely on visible impact metrics as proxies for partner quality \citep{podolny2001networks, connelly2011signaling}, while communication conditions influence the returns to collaboration \citep{olson2000distance, cummings2007coordination}.

\section{Literature Review}

\subsection{International Research Collaboration Networks}

By the turn of the 21st century research teams rather than lone scientists were responsible for the production of a large proportion of new research \citep{katz1997research, fortunato2018science}. These teams were increasingly composed of researchers hailing from a diverse mix of countries from around the world \citep{wagner2005network}. Accordingly, international research collaboration (IRC), defined as instances where published research includes two or more authors with distinct country affiliations \citep{frame1979international}, rapidly became a norm in the global corpus of science and technology \citep{chen2019international}. IRC is now an important priority for policy makers in most countries around the world \citep{wagner2015continuing}.

Although research collaboration can be traced back to the process of scientific professionalization that developed in France during the 17th to 19th centuries \citep{beaver1978studies, beaver1979studies}, a coherent body of research on collaboration did not begin to accumulate until the mid-20th century \citep{katz1997research}. Collaboration involving international teams emerged as a topic of concern shortly thereafter during its latter half. \citet{frame1979international} conducted one of the first bibliometric studies of IRC noting several antecedents, such as small country size, basic versus applied research orientation, and geographic proximity. \citet{luukkonen1992understanding} noted the effect of travel, communication technology, and government programs on IRC. The literature continued to develop alongside the expansion of IRC as a practice in science and technology \citep{luukkonen1992understanding}. Coinciding with the emergence and rise of the Internet, the 1990s to 2000s witnessed a sharp increase that has continued ever since \citep{wagner2005network, chen2019international}. Since the early 21st century our understanding of IRC patterns, drivers, and consequences have only grown, yet important questions remain, as do opportunities for generating new insights in the context of massive data sets and dramatically improved computational capabilities. 

While instances can be found in earlier literature of the treatment of IRC as a network object \citep{schubert1990international, melin1996studying}, the development of the topic through the application of the tools of network science remains a relatively niche area of research \citep{chen2019international}. Early studies of scientific collaboration networks sought to uncover complex processes that stimulate self-organization, such as preferential attachment, occurring in a variety of network types across subject domains \citep{newman2001structure, newman2004coauthorship}. Through this lens, the IRC network resembles a dynamic self-organizing entity, i.e., complex adaptive system \citep{katz2016complex}. As such, the IRC network is characterized by decentralized co-author level interactions that produce emergent patterns and are not dictated top-down by central authorities \citep{wagner2005network}, bracketing for the moment anomalies such as China \citep{wagner2024science, zhang2022exploring}. 

However, while self-organizing systems lack central coordination, they can still be significantly affected by institutional rules, norms, and factors of political governance. Self-organization, as a general process in complex systems, does not nullify the influence of national systems of science and innovation \citep{freeman1992}. For example, recent work has demonstrated the effects of national governance on the formation of the IRC network \citep{whetsell2023democratic, whetsell2024academic}. While the relative merits of taking the view of the global network versus the national system may be debated, it is useful to think of these as interconnected macro-micro layers in a more general system \citep{coleman1990foundations}.

A relatively recent systematic review of IRC revealed that network-based studies remain relatively scant \citep{chen2019international}. Further, many studies that do leverage network analysis in IRC research focus on describing the international landscape of collaboration \citep{schubert1990international, melin1996studying, glanzel2004analysing, niu2014network, vakilian2015bibliometric, mallik2014bibliometric, fu2022evolving} and/or focus on case studies of specific disciplines \citep{nita2019empowering, kuzhabekova2015mapping, fu2022evolving}. There remains a lack of large scale inferential work testing the effects of IRC networks on outcomes of interest such as the performance of research.

\subsection{Network Effects on Performance}

In research evaluation, performance is generically defined as the influence or impact of a publication and is usually operationalized through citation-based indicators \citep{garfield1979citation, sinatra2016quantifying, aksnes2019citations}. At the national level, performance is the impact of the research attributed to authors affiliated with the country \citep{king2004scientific}. Recent innovations have produced a performance indicator that can be used at the country level known as the fractional field-weighted citation impact (FWCI) \citep{purkayastha2019comparison, wagner2018openness}. Further, the term \textit{performance} is also used because a large body network-based research, particularly in organization theory, has examined the effects of network position on performance metrics in a variety of settings \citep{powell1999network, zaheer2005benefiting, koka2008designing}.

Extensive evidence suggests that international papers tend to have greater citation impact than those produced domestically, resulting from both functional integration and audience effects \citep{narin1991scientific, glanzel2002distributional, glanzel2001national, persson2010highly, wagner2017growth, kwiek2021large}. Research suggests that collaboration facilitates the functional integration of information and resources \citep{Wray_2002}, and also produces broader audience effects \citep{wagner2019international, thelwall2023coauthored}, which both lead to higher citation performance. While \citet{adams2024national} argue that rising national citation averages reflect the growing share of collaborative papers rather than gains in performance per paper, our use of fractional attribution addresses this counting-method concern directly \citep{waltman2015field}. For junior scholars, even indirect international collaboration ties through domestic collaborators with global connections has been shown to lead to enhanced access to novel research opportunities, increasing research productivity and citation impact \citep{chen2025benefits}. However, some research suggests that even though higher IRC may lead to higher citation impact, IRC may actually produce more conventional research \citep{wagner2019international}, often failing to achieve atypical ``hit'' paper status identified by \citet{uzzi2013atypical}.

Far less research has been conducted that explicitly leverages the tools of network science to model relational IRC data as a formal network object. Viewing IRC through the lens of complexity and network science may reveal key insights about its potential effects on research performance \citep{wagner2005network, katz2016complex, de2024empirical}. Research tends to focus on authors, papers, or countries as independent observations and does not account for inter-dependencies that arise in complex systems. Even many network studies of IRC still do not account for complex dependencies, instead leveraging the network approach for its descriptive utility in mapping and ranking countries. Analysis of the IRC network can not only produce descriptive insights but can also reveal fundamental antecedents and practical consequences at a global scale. Rather than using standard linear models to estimate effects of country count on citation impact, inferential network models can account for complex dependencies in relational data, such as preferential attachment, transitivity, homophily, etc. in the estimation of IRC effects on performance \citep{cranmer2020inferential}. 

Research collaboration ties represent the potential for exchange of information and resources between researchers \citep{katz1997research, Wray_2002}. Occupying central positions in social networks often confers advantages related to access, status, and influence over those that occupy positions at the periphery or those that are network isolates \citep{granovetter1985economic, powell1999network, podolny2001networks}. Thus, researchers who occupy favorable network positions within the aggregate network structure may exploit the advantages associated with access to information and resources that may be leveraged toward the production of high performing research. 

A number of articles have leveraged the tools of network analysis to estimate network effects on research performance in small samples, specific disciplines, or in co-author networks without a focus on the international element. \citet{wang2015characteristics} show that international collaboration is associated with enhanced citation impact in a sports sciences co-author network. \citet{badar_knowledge_2015} model a coauthor network of Pakistani researchers showing the positive association between degree centrality and betweenness centrality with research performance in chemistry. \citet{li2013co} show significant positive effects of betweenness centrality on citation impact through the lens of social capital theory in an information systems research co-author network. \citet{fares2021stakeholder} show that network variables such as degree and betweenness centrality, structural holes, and tie strength have become increasingly correlated with citation impact over the last three decades in the sub-field of stakeholder theory and management. Recently, \cite{li2022untangling} analyze collaboration networks as a form of social capital, finding that scientists' coauthor quality explains a substantial portion of productivity and prominence differences.

Despite advances in the field, there are a number of limitations to the extant literature testing IRC network effects on research performance. Studies tend to focus on one or a few disciplines, rely on small samples, make inconsistent use of longitudinal data to test inherently dynamic hypotheses, and depend on standard linear models that cannot accommodate the interdependence of countries embedded in a shared global network.

\begin{quote}
\textbf{Hypothesis 1:} \textit{Increasing IRC network centrality enhances national S\&T research performance.}
\end{quote}

Following the main hypotheses, auxiliary working hypotheses are derived which motivate further empirical testing. The relationship between the international research collaboration (IRC) network and national S\&T performance is thought to be shaped by important contextual moderators. First, geographical proximity plays an important role in learning \citep{boschma2005proximity} and is known to affect collaborative tie formation, knowledge exchange, and spillovers \citep{katz1994geographical, olson2000distance, cummings2007coordination}. A relatively large literature shows that geographic distance impedes knowledge diffusion and international research collaboration across boundaries \citep{frame1979international, hoekman2010research, waltman2011globalisation}. Despite the development of communication technologies, national citation networks continue to reflect localized knowledge spillovers, suggesting that scientists are more likely to cite papers within the same country or close region \citep{abramo2020role,abramo2020knowledge, fernandez2016proximity, wuestman2019geography}. Geographic distance limits the chances of serendipitous encounters or informal relationship building, which are essential for effective collaboration, and tacit and contextual knowledge are more difficult to share remotely in the absence of face-to-face interaction \citep{abramo2020role}. 

While several studies have examined the direct effect of distance on collaboration propensity and citations, to our knowledge, none have used distance as a moderator of the effect of research collaboration on research performance. Thus, we advance the following working hypothesis as an auxiliary to the first primary hypothesis. 

\begin{quote}
\textbf{Hypothesis 1a:} \textit{The positive effect of IRC network centrality on national S\&T
research performance is weaker for geographically remote countries.}
\end{quote}

A second auxiliary working hypothesis concerns the moderating effect of language. Common language lowers communication costs and is a critical factor in effective international exchange \citep{melitz2014native}. Language differences have been shown to hinder effective international research collaboration \citep{luukkonen1992understanding, hoekman2010research}. Linguistic distance often not only leads to misinterpretation but also causes negative emotional experiences and a sense of marginalization among non-native speakers \citep{tenzer2013leading, hwang2005inferior}, undermining mutual understanding and richer academic discussion. These barriers reduce both the frequency and quality of scholarly discussion, limiting collaborators' ability to share nuanced ideas regardless of their expertise \citep{hwang2013effects}. A shared language reduces the risk of misunderstanding and enables clear communication. It also fosters interpersonal working relationships and the efficient exchange of nuanced and complex knowledge between collaborators. Thus, while common language between collaborators is neither a necessary nor a sufficient condition, it nevertheless facilitates effective international research collaboration \citep{hou2021impact}. 

Similar to geographic distance, while several studies have examined the direct effect of shared language and linguistic proximity on collaboration and citations, we are aware of none that have used language as a moderator of the effect. Thus, we advance the following working hypothesis as the second auxiliary to the first primary hypothesis. 

\begin{quote}
\textbf{Hypothesis 1b:} \textit{The positive effect of IRC network centrality on national S\&T research performance is stronger for countries with greater linguistic reach.}
\end{quote}

\subsection{Performance Effects on Networks}

While many studies have emphasized the functional effects of research collaboration on performance, fewer studies have examined the reverse pathway, in which performance feeds back into the social processes that drive network evolution. Preferential attachment is a general social mechanism that has gained virtual law-like status in studies of social networks \citep{newman2003structure, barabasi2002evolution}. It operates by generating increasing returns to popularity as network entrants seek out well connected alters. Preferential attachment is one of a range of phenomena that operate on the basis of cumulative advantage often manifesting in power-law like data distributions and is also related to Zipf and Pareto distributions, as well as Bradford's and Lotka's Laws \citep{price1965networks}. In the context of wealth generation, the effect is more colloquially known as the Matthew Effect, i.e. ``the rich get riche'' phenomenon. \citet{merton1968matthew} applied the Matthew effect to the behavior of researchers to explain how high status scientists tend to generate increasing popularity and, as a secondary consequence, reduce the visibility of others. 

In the IRC context, preferential attachment is a mechanism through which new researchers entering the network tend to seek out the most well-connected and highest performing researchers in the network \citep{newman2001structure, wagner2005network}. Research performance, often indicated by citation impact, is a highly visible metric on which other researchers optimize their searching strategies when selecting new collaboration partners. However, returns to collaborations with ``star'' scientists are not uniform. Recent evidence indicates that high performing junior scholars are more likely to benefit from research collaboration as they may be in a better position to leverage collaborative opportunities toward research performance \citep{liu2022scientific}. This suggests that the early performance advantage of collaborating with high performers may be amplified over time.

This logic also applies to various related areas. Research funding often relies on the past performance of grantees, since prior grant performance increases the chance new grant proposal acquisition from internal and external funding sources \citep{auranen2010university}. There is also some evidence that peer review is more favorable to prestigious authors in the context of a single-blind review process which the majority of science journals use \citep{tomkins2017reviewer}, and hence might be attractive as collaborators who assume the corresponding author role. Finally, similar to the cumulative network effect of co-authorship, it is also well established that papers and authors with higher citation counts are more likely to be cited in the future, tending to also display power law distributions \citep{redner1998popular} indicative of preferential attachment \citep{jeong2003measuring} and hence the Matthew effect \citep{perc2014matthew}. 

Further, there is some evidence that high performers also tend to seek out collaborative opportunities with other high performers through a process known as assortative mixing \citep{newman2002assortative}. This social process is closely related to preferential attachment, but is also similar to the well known dynamic of homophily, illustrated colloquially as ``birds of a feather flock together". In this sense, high performing, and hence high status, researchers tend to seek each other out as partners, which may further generate future performance enhancements and hence greater returns to collaboration.

However, collaborator selection and formation do not merely obey social mechanisms. They also are driven in part by functional needs related to combining knowledge, skills, and resources \citep{Wray_2002,jha2010relational}. In other words, researchers often seek to be connected to 'star' alters in order to gain access to funding, equipment, knowledge, and skills related to the functional requirements of producing cutting edge research \citep{yadav2023does,leahey_sole_2016, davies2022research}. Lastly, \citet{abramo2009research} and \citet{abramo2011relationship} pose the question of whether there is some evidence of a virtuous cycle between performance and the IRC network.

This prior body of research suggests the following primary hypothesis that serves as the second leg of the core proposition about co-evolution. While hypothesis one identifies expectations about network effects on performance, hypothesis two specifies expectations about the social effect of performance on collaboration.  

\begin{quote}
\textbf{Hypothesis 2:} \textit{Increasing S\&T research performance enhances IRC network evolution.}
\end{quote}

Two auxiliary working hypotheses are elaborated by this hypothesis. The effect of national research performance on the evolution of international research collaboration (IRC) networks may be moderated by both geographic distance and language proximity in important ways.

Although research suggests that geographical distance may discourage scientists from working with distant collaborators, some scholarship also suggests that star scientists are more likely to overcome, or even leverage, geographical distance in research collaborations \citep{abramo2019collaboration}. 

We suspect that researchers face information asymmetries when selecting partners for collaboration in ways that can reinforce performance effects on the network through signaling mechanisms. Selecting a collaborator is a decision under uncertainty: a prospective partner's true research capability is not directly observable, particularly across national and linguistic boundaries. Under such information asymmetry, actors rely on conspicuous signals in order to avoid suboptimal outcomes \citep{spence1978job, connelly2011signaling}. The weight placed on these signals rises as uncertainty about partner quality increases \citep{podolny2001networks}. 

Given the prospect of engaging in a high distance research collaboration, researchers may lack information about their potential collaborator's capabilities, expertise, or personality. In such cases, observable signals, such as publication counts, citation counts, awards, or institutional ranking, become critical for estimating future benefits \citep{abramo2024moderating, onuchic2023signaling}. This signaling mechanism may be particularly important for newcomers or peripheral actors in international scientific networks, who often lack established ties or local knowledge \citep{ozmel2013signals}. These mechanisms operate at the level of the individual researcher, and the hypothesis tests are an expression of that logic aggregated at the country level. Thus, we advance the following auxiliary working hypothesis that leverages geographic distance as a magnifier of the performance to collaboration relationship.

\begin{quote}
\textbf{Hypothesis 2a:} \textit{The positive effect of national S\&T research performance on IRC network evolution is stronger when collaborators are geographically distant.}
\end{quote}

While geographic distance is expected to magnify the performance effect on tie formation (H2a), a common or shared language may work in the opposite direction. Shared language reduces the information asymmetries that give performance signals their value \citep{spence1978job, connelly2011signaling}. When collaborators share a language, they can interact more easily and assess one another's capabilities directly, rather than relying on formal indicators such as citation-based performance metrics \citep{melitz2014native}. Because reliance on visible signals rises with uncertainty \citep{podolny2001networks}, it should weaken the effect of citation performance on selection as shared language reduces uncertainty about a collaborator's quality. It should be noted that a shared language could also work in the opposite direction, since high performing countries are more visible within their own linguistic communities. Nevertheless, we expect the signaling effect of research performance on tie formation to weaken when potential collaborators share a common language. Thus, we advance the following auxiliary working hypothesis that leverages common language as an attenuator of the performance to collaboration relationship.

\begin{quote}
\textbf{Hypothesis 2b:} \textit{The positive effect of national S\&T research performance on IRC network evolution is weaker when collaborators share a common official or primary language.}
\end{quote}

\section{Methods}

In recent decades, network methodologists have developed novel statistical tools for the analysis of longitudinal network data known as stochastic actor-oriented models (SAOM) \citep{snijders2001statistical,snijders2017modeling}. These models can estimate the effects of actor-behavior on network dynamics (selection effects) while also accounting for endogenous dependencies in network data \citep{cranmer2020inferential}. Actor 'behavior' refers to node attributes, such as country GDP, population size, and research performance. Network dynamics refers to tie formation over time contributing to network evolution. Extensions of the SAOM, known as co-evolution models, enable modeling reciprocal effects of network dynamics on actor behavior (influence effects) \citep{snijders2017modeling}. 

SAOMs are implemented in the \texttt{R} package \texttt{RSiena}, which stands for Simulation Investigation for Empirical Network Analysis \citep{snijders2025manual}. In contrast to the traditional statistical regression framework, SAOMs are useful for analyzing complex dependencies in longitudinal network data and also enable simultaneous analysis of two-way reciprocal effects. SAOM is suitable for testing our hypotheses since it can separate selection and influence effects and identify whether there is a reinforcing feedback loop between different effects. In the present study, selection captures how countries form ties with others based on performance levels, whereas influence captures how network dynamics affect research performance. The actors in the model are the 172 countries. Tie changes and performance changes are therefore modeled at the country level. The individual-level theoretical mechanisms motivate these country-level predictions rather than being estimated directly, following the macro-micro logic described by \citet{coleman1990foundations}.

SAOM assumes that actors make choices to change ties leading to network changes. Given the opportunity for change, the probability that actor $i$ moves from the current state of the network $x^0$ to the next state $x$, conditional on the current behavior $z$, follows a multinomial logit form, represented by Equation~(\ref{probchange_net}) (as described in the \texttt{RSiena} manual \citep{snijders2025manual}).

\begin{equation}
\label{probchange_net}
    P(X_{\text{next}} = x \mid X_{\text{current}} = x^0,\; Z = z,\; \text{actor } i) =
    \frac{\exp\left(f_i^{\text{net}}(x, z)\right)}
    {\displaystyle\sum_{x' \in C_i^{\text{net}}} \exp\left(f_i^{\text{net}}(x', z)\right)}
\end{equation}

\noindent Let $X$ denote the network and $Z$ the behavior variable. Lower case $x$ and $z$ denote the values these variables take. Here $C_i^{\text{net}}$ is the set of network states obtainable from $x^0$ when actor $i$ changes at most one of its ties. This set includes the current state $x^0$, so leaving the network unchanged is one of the available choices. These expressions present the standard form of the model. In the non-directed variant used here, the micro-step is adapted so that tie changes proposed by one actor may also involve the other actor, as described below. The network evaluation function is a linear combination of effects, represented by Equation~(\ref{evalfunc_net}).

\begin{equation}
\label{evalfunc_net}
    f_i^{\text{net}}(x, z) = \sum_k \beta_k^{\text{net}}\, s_{ik}^{\text{net}}(x, z)
\end{equation}

\noindent where $\beta_k^{\text{net}}$ are the network parameters and $s_{ik}^{\text{net}}(x, z)$ are the network effect statistics that may depend on both the network and behavior. Selection effects, such as homophily, are captured by statistics that incorporate the behavioral variable into tie formation decisions. 

Analogously, given the opportunity for change, the probability that actor $i$ changes behavior from $z^0_i$ to $z_i$, conditional on the current network $x$, is given by Equation~(\ref{probchange_beh}).

\begin{equation}
\label{probchange_beh}
    P(Z_{i,\text{next}} = z_i \mid Z_{i,\text{current}} = z^0_i,\; X = x,\; \text{actor } i) =
    \frac{\exp\left(f_i^{\text{beh}}(x, z_i)\right)}
    {\displaystyle\sum_{z'_i \in C_i^{\text{beh}}} \exp\left(f_i^{\text{beh}}(x, z'_i)\right)}
\end{equation}

\noindent where $C_i^{\text{beh}}$ is the set of behavior values obtainable from $z^0_i$ by a change of at most one unit, that is $\{z^0_i - 1,\ z^0_i,\ z^0_i + 1\}$ truncated at the ends of the ordinal scale. The evaluation function is computed on the behavior vector in which actor $i$'s value is replaced by the candidate value and all other values are unchanged.

The behavioral evaluation function combines behavioral tendencies and network-based effects to model how actors evaluate potential behavioral changes, represented by Equation~(\ref{evalfunc_beh}).

\begin{equation}
\label{evalfunc_beh}
    f_i^{\text{beh}}(x, z) = \sum_k \beta_k^{\text{beh}}\, s_{ik}^{\text{beh}}(x, z)
\end{equation}

\noindent where $\beta_k^{\text{beh}}$ are the behavioral parameters and $s_{ik}^{\text{beh}}(x, z)$ are the behavioral effect statistics. Influence effects, such as the average behavior of an actor's network neighbors, are captured by statistics that model how network ties affect behavioral change. 

The rate function $\rho_m^{\text{net}}$ controls the frequency of tie change opportunities for each actor in period $m$ between consecutive observations, while the behavioral rate function $\rho_m^{\text{beh}}$ controls the frequency of behavioral change opportunities in the same way.

Parameters are estimated with the simulation-based method of moments implemented in \texttt{RSiena} \citep{snijders2001statistical}. The algorithm uses stochastic approximation to find parameter values where simulated statistics match their observed targets. Convergence is assessed from the $t$ ratios for deviations from targets and from the overall maximum convergence ratio reported for each model.

The SAOM's treatment of time addresses simultaneity between collaboration and citation impact. The model is not a contemporaneous regression of performance on network position. Rather, an unobserved continuous-time process between observations is posited in which actors take sequential micro-steps, altering a single network tie or changing performance by one ordinal level. Each step is conditional on the current state of both the network and behavior. Network change at any instant depends on the current state of performance (selection), while performance change depends on the current state of the network (influence). As such, the separation of these reciprocal bi-directional relations is achieved by the model, rather than by lagging observed variables.

The co-evolution SAOM is the only established statistical framework that can jointly estimate the reciprocal influence of both a) the network on actor behavior and b) actor behavior on the network \textit{longitudinally}, while also accounting for endogenous structural dependencies such as preferential attachment and transitivity. The SAOM is thus the most appropriate model for testing the core proposition regarding reciprocal effects of network selection and performance influence.

\subsection{Data \& Variables}

Data are compiled from numerous sources, including Clarivate's Web of Science, Elsevier citation metrics, the World Bank's economic indicators, the KOF's globalization metric, and the Varieties of Democracy Project's democracy indicator. The variable selection strategy prioritized accounting for both omitted variable bias and also for maximizing data coverage to avoid country omission.

The IRC networks are generated by parsing the Web of Science XML database from 1993 to 2022. Each record is parsed in \texttt{Python} and filtered to the science and technology subject heading. Before any ties are counted, country labels in each record are harmonized to ISO3 codes using a static lookup table applied at the record level. This ordering is deliberate. Disambiguating to a single ISO3 code before counting prevents a small amount of inflation that results from merging variants after co-occurrences are counted. A general-purpose reference resolver (the \texttt{country\_converter} package; \citealp{stadler2017country}) covers standard name variation but not the WoS-specific country naming ambiguities, which include truncations from legacy records, misspellings, and occasional non-country substitutes such as city names. The lookup table is WoS-derived for this reason. It was assembled earlier with the assistance of a large language model, and every mapping occurring in the data has since been checked against the reference resolver. Historical entities without a single successor state, codes for entities without continuous coverage across the study period, and unresolvable labels were excluded. From the harmonized records we construct thirty one-mode, symmetric, valued matrices, one per year. The rows and columns are country ISO codes and the cell values are non-negative integers. Each record contributes at most one to a country pair, so a cell value is the number of records in that year on which the two countries co-occur in the author affiliations.

The SAOM model imposes two important data requirements. Network ties must be binary because the actor-oriented micro-step defining the SAOM toggles the dyadic tie on or off. As such, the approach taken here is to transform the valued country-country matrices into binary form using the disparity filter \citep{serrano2009extracting}, a principled node-level backbone-extraction method that retains only ties carrying a disproportionate share of a node's total tie strength relative to a null model ($\alpha = 0.05$). This procedure is implemented using the \texttt{Backbone} package \citep{neal2014backbone, neal2022backbone}. Unlike a single global threshold, the disparity filter evaluates each tie against its own node's baseline, preserving locally significant structure but changing the interpretation of a tie from simple co-authorship to statistically significant ``backbone'' collaboration. Because binarization reshapes the degree distribution for degree-based effects and the default $\alpha = 0.05$ removes the large majority of raw co-authorship ties, a sensitivity check was conducted using a less restrictive threshold that preserves more ties ($\alpha = 0.10$; Appendix~\ref{app:robustness}, Model~A2). Because the \texttt{Backbone} procedure is conducted yearly, temporal variation is created across the network panel suitable for analysis by \texttt{RSiena}. 

The other constraint of the standard co-evolution model is that the behavioral dependent variable must be ordinal with a limited number of categories (usually 10 or under), because behavioral change is likewise modeled as unit micro-steps along an ordinal scale. A recent development in the SAOM framework which implements continuous behavioral variables \citep{niezink2019discrete} was attempted but failed to converge. We therefore discretized FWCI into ten ordinal levels using decile breaks computed on the pooled country-year distribution, so that a given level corresponds to the same absolute FWCI range in every wave. A small number of country-years with no performance data (34 of 5,160) were retained as missing and handled by \texttt{RSiena}'s internal missing-data procedures \citep{huisman2008treatment}. Lastly, since co-authorship ties are inherently symmetric, the network was modeled as non-directed using \texttt{RSiena}'s initiative model for non-directed networks (modelType~3). 

The measure of national research performance used here is Elsevier's fractional field-weighted citation impact (FWCI) \citep{purkayastha2019comparison}. The FWCI approach divides a paper's citation count by the expected count for its field, publication year, and document type so a score reflects impact rather than field or age \citep{purkayastha2019comparison}. At the country-year level FWCI provides an aggregate score by averaging over all individual paper FWCI scores for the country. Fractional counting of citations is a separate improvement that has become a standard approach in bibliometrics that disentangles attribution problems in coauthored papers \citep{waltman2015field} by splitting each paper into contributor shares that sum to one, in this case to avoid inflating a country's citations from co-authored papers. The data were provided through email contacts at Elsevier as a single data sheet in long panel format for all-fields and including a breakdown of 6 fields. Three of these categories approximately correspond to WoS's science and technology heading. Therefore, the median of these values was taken across these three categories for each country-year in order to match with the WoS generated IRC network year. \texttt{RSiena} requires actor-behavior data formatted in wide-panel format. One concern with citations is that they accrue over multiple years. However, this is also addressed by the FWCI approach, which is computed over a multi-year citation window anchored to publication year, so the measure already incorporates delayed citation accrual rather than reflecting same-year citations alone.

To prevent biasing the analysis by focusing on developed national science systems and to preserve the largest possible sample size of countries, we favor country covariates with broad data coverage across time. Economic and demographic covariates were selected from World Bank data, accessed through the \texttt{R} package \texttt{WDI}, and include GDP per capita, population size, and urbanization. The KOF Globalisation Index \citep{dreher2006does, gygli2019kof} is also used, accessed through the Quality of Government Standard dataset \citep{teorell2025qog}. This index captures a country's economic integration with the world and has broad coverage across the study period. Publication volume is also included, measured as each country's fractional science and technology publication count from Web of Science and log-transformed, so that the effects of network centrality and performance can be distinguished from sheer research output. More fine-grained measures of research investment such as R\&D expenditure (GERD or RDGDP) have narrower country coverage than GDP per capita and are therefore not included in the main specification. Instead, we test R\&D intensity as a substitute for GDP per capita in a robustness model (Appendix~\ref{app:robustness}, Model~A1). To account not only for economic and population based covariates but also for governance factors known to affect IRC and citation impact \citep{whetsell2023democratic, whetsell2021democracy}, a variable that accounts for democracy (polyarchy) was collected from the Varieties of Democracy Project accessed through the \texttt{R} package \texttt{VDEM} \citep{VDEM2024}. This data set also has broad coverage across time. 

Two dyadic covariates were included. Geographic distance has a well-known negative effect on IRC. Thus, a geographic distance matrix is included as an edge covariate. The data is unvarying over time, which may prove unrealistic in some cases if country borders change significantly during the time frame. In addition, to account for common culture we also include shared official or primary language between collaborators on international collaboration \citep{luukkonen1993measurement}. International collaboration can be discouraged by language barriers \citep{hwang2013effects, tenzer2014impact}, but can also be facilitated when collaborators share a common language due to ease of communication \citep{hou2021impact, melitz2014native}. Geographic distance and common official or primary language were collected from the Centre for Prospective Studies and International Information (CEPII) and converted into matrix format. 

\begin{table}[H]
\centering
\footnotesize
\setlength{\tabcolsep}{4pt}
\renewcommand{\arraystretch}{0.9}
\caption{Description of Variables for IRC Network Evolution}
\label{vardesc}
\begin{tabular}{l p{0.68\textwidth}}
Variable Name & Description of Variable and Model Term \\
\hline
\multicolumn{2}{c}{Effects on Network Dynamics (Selection)}\\
\hline
Intercept & Base-line tie formation tendency (network density).\\
Transitivity (gwesp) & Endogenous triadic closure (friend-of-a-friend effect); geometrically weighted edgewise shared partners.\\
Pref. Attach (degPlus) & Endogenous preferential attachment (Matthew effect); node degree plus.\\
Dist (X) & Geographic distance between countries (km/1000, log-transformed); constant dyadic covariate.\\
Lang (X) & Shared official or primary language between countries; constant dyadic covariate.\\
FWCI (egoPlusAltX) & Direct effect of national FWCI (research performance).\\
FWCI (simX) & Similarity in FWCI between countries (homophily).\\
Poly (egoPlusAltX) & Direct effect of democratic polyarchy.\\
Poly (simX) & Similarity in democratic polyarchy between countries.\\
Glob (egoPlusAltX) & Direct effect of globalization.\\
Glob (simX) & Similarity in globalization between countries.\\
GDP (egoPlusAltX) & Direct effect of GDP per capita (log-transformed).\\
GDP (simX) & Similarity in GDP per capita between countries (log-transformed).\\
Pop (egoPlusAltX) & Direct effect of population size (log-transformed).\\
Pop (simX) & Similarity in population size between countries (log-transformed).\\
Urb (egoPlusAltX) & Direct effect of urbanization.\\
Urb (simX) & Similarity in urbanization between countries.\\
Pubs (egoPlusAltX) & Direct effect of fractional S\&T publication volume (log-transformed).\\
Pubs (simX) & Similarity in publication volume between countries.\\
Dist X FWCI (egoPlusAltX) & Interaction of distance with national FWCI.\\
Lang X FWCI (egoPlusAltX) & Interaction of shared language with national FWCI.\\
\hline
\multicolumn{2}{c}{Effects on Performance (Influence)}\\
\hline
Linear shape & Behavior-change intercept; linear tendency in performance over time.\\
Quadratic shape & Quadratic tendency for performance change.\\
IRC Net. Deg. (outdeg) & Effect of network degree (centrality) on performance.\\
IRC Avg. Alter (avAlt) & Effect of average partner performance (peer influence).\\
Dist (effFrom) & Effect of geographic remoteness (mean distance to all countries) on performance change.\\
Lang (effFrom) & Effect of linguistic reach (mean shared-language incidence) on performance change.\\
Poly (effFrom) & Effect of democratic polyarchy on performance change.\\
Glob (effFrom) & Effect of globalization on performance change.\\
GDP (effFrom) & Effect of GDP per capita on performance change.\\
Pop (effFrom) & Effect of population size on performance change.\\
Urb (effFrom) & Effect of urbanization on performance change.\\
Pubs (effFrom) & Effect of publication volume on performance change.\\
Deg X Dist (effFrom) & Interaction of network degree with geographic remoteness\\
Deg X Lang (effFrom) & Interaction of network degree with linguistic reach\\
\hline
\end{tabular}
\end{table}

After processing the data in light of all variables used in the models, the final sample included 172 countries. \texttt{RSiena} models requires uniform node sets across years. Therefore, the complete 172 node set was imposed on the matrix structure across the time period. Table~\ref{vardesc} provides a description of all variables used in the \texttt{RSiena} models. The table is divided into two parts corresponding to both sides of the co-evolution model, the first containing selection effects of country attributes and other endogenous processes on network tie formation and evolution, the second containing influence effects of network properties and processes on behavior operationalized as FWCI research performance. The \texttt{RSiena} model term is also provided in parentheses next to the variable name.

\section{Results}\label{sec:results}

The results of the hypothesis tests are presented in Table~\ref{tab:Co-Evolution Models} which shows five co-evolution stochastic actor-oriented models. The top half of the table shows the effects of exogenous node attributes and endogenous processes on network formation and evolution, while the bottom half shows the effects of the network on scientific performance measured with fractional field-weighted citation impact (FWCI). For brevity, we term the top half the effects on the network (or network side), which are relevant to the H2 tests, and the bottom half the effects on performance (or performance side), which are relevant to the H1 tests. These are also often commonly referred to as selection and influence effects respectively. The presentation strategy is to incrementally increase the complexity of the models in order to explore the effects of additional variables, where the first model shows a very minimal model and the fifth model includes the complete list of relevant variables. 

The first model is a minimalist setup with only an intercept, the FWCI main effect (egoPlusAltX), and homophily effect (simX) on the network side; and the linear and quadratic shape terms on the performance side. The main effect of FWCI on the network is positive and significant while the homophily effect is negative, meaning countries with higher performance tend to generate more backbone ties over time, while at the same time countries are more likely to form ties with others of dissimilar performance levels. This is a fairly consistent observation across the study and provides support for H2 regarding the social effect of performance on backbone tie formation and network evolution. On the performance side of the model, the linear and quadratic shape terms indicate whether there is a general change in the overall FWCI over time. However, since FWCI is normalized around one, at the country level, linear or quadratic shape indicates whether countries overall are converging or diverging from the global average. In the first model these terms are positive with mixed significance. 

In model two on the network side, endogenous network terms are added representing preferential attachment and transitivity processes. These processes are well established in the literature and have a strong significant effect on network evolution. One immediate result of including these two endogenous network controls is that the magnitude of the main FWCI effect is dramatically reduced, and the homophily effect of FWCI is rendered insignificant. This immediately called into question the support for H2. On the performance side of the second model, the network degree effect of the country is added, showing a positive and significant effect, indicating that higher country network centrality is associated with stronger performance, providing support for H1. In addition, we included the average-alter term, which captures the mean performance level of a country's network partners. Its positive and significant coefficient indicates that countries whose collaborators perform well tend to improve their own performance, which indicates a peer-influence (social contagion) effect operating alongside the direct benefit of connectivity.  

In the third model on the network side, we add two constant dyadic covariates, one for geographical distance and one for common language. As anticipated, the first is negative and the second is positive. Interestingly, the main effect of FWCI appears to increase in magnitude when adding these terms. In the model specification process, this observation indicated there may be an interaction effect at play. On the performance side of the model three we similarly added terms for distance and language. However, because this side of the model is network effects on performance, the variables are mean aggregated at the country level rather than the dyadic level. 

\begin{table}[H]
\centering
\small
\caption{Co-Evolution Models}
\label{tab:Co-Evolution Models}
\adjustbox{max width=\textwidth}{%
\begin{tabular}{lccccc}
Variable & Model 1 & Model 2 & Model 3 & Model 4 & Model 5 \\
\hline
\multicolumn{6}{c}{Effects on Network Dynamics (Selection)} \\
\hline
Intercept & -0.969*** (0.014) & -2.630*** (0.038) & -2.970*** (0.043) & -3.000*** (0.045) & -3.306*** (0.059) \\
Transitivity (gwesp) &  & 0.675*** (0.024) & 0.608*** (0.025) & 0.611*** (0.026) & 0.384*** (0.034) \\
Pref. Attach (degPlus) &  & 0.016*** (0.000) & 0.019*** (0.000) & 0.018*** (0.000) & 0.009*** (0.000) \\
Dist (X) &  &  & -0.692*** (0.016) & -0.742*** (0.019) & -1.000*** (0.023) \\
Lang (X) &  &  & 0.679*** (0.023) & 0.664*** (0.026) & 1.059*** (0.030) \\
FWCI (egoPlusAltX) & 0.154*** (0.004) & 0.018*** (0.004) & 0.033*** (0.004) & 0.045*** (0.005) & 0.066*** (0.007) \\
FWCI (simX) & -0.524*** (0.051) & 0.015ns (0.052) & -0.073ns (0.053) & -0.128* (0.056) & -0.157* (0.069) \\
Poly (egoPlusAltX) &  &  &  &  & 0.315*** (0.036) \\
Poly (simX) &  &  &  &  & 0.333*** (0.048) \\
Glob (egoPlusAltX) &  &  &  &  & -0.001ns (0.001) \\
Glob (simX) &  &  &  &  & 0.567*** (0.072) \\
GDP (egoPlusAltX) &  &  &  &  & 0.052*** (0.015) \\
GDP (simX) &  &  &  &  & 0.079ns (0.110) \\
Pop (egoPlusAltX) &  &  &  &  & 0.185*** (0.011) \\
Pop (simX) &  &  &  &  & 0.572*** (0.082) \\
Urb (egoPlusAltX) &  &  &  &  & 0.001* (0.001) \\
Urb (simX) &  &  &  &  & 0.027ns (0.073) \\
Pubs (egoPlusAltX) &  &  &  &  & 0.136*** (0.009) \\
Pubs (simX) &  &  &  &  & -1.672*** (0.101) \\
Dist x FWCI (egoPlusAltX) &  &  &  & 0.021*** (0.005) & 0.046*** (0.005) \\
Lang x FWCI (egoPlusAltX) &  &  &  & 0.002ns (0.006) & -0.023*** (0.007) \\
\hline
\multicolumn{6}{c}{Effects on Performance (Influence)} \\
\hline
Linear shape & 0.016ns (0.008) & -0.295*** (0.029) & -0.292*** (0.029) & -0.298*** (0.029) & -0.267*** (0.033) \\
Quad. shape & 0.012*** (0.002) & -0.005* (0.002) & -0.005* (0.002) & -0.006** (0.002) & -0.013*** (0.002) \\
IRC Net. Deg. (outdeg) &  & 0.014*** (0.001) & 0.015*** (0.001) & 0.015*** (0.001) & 0.013*** (0.001) \\
IRC Avg. Alter (avAlt) &  & 0.081*** (0.010) & 0.077*** (0.010) & 0.080*** (0.010) & 0.075*** (0.010) \\
Dist (effFrom) &  &  & 0.045ns (0.032) & 0.128** (0.042) & 0.098* (0.043) \\
Lang (effFrom) &  &  & 0.213** (0.074) & 0.011ns (0.094) & 0.071ns (0.099) \\
Poly (effFrom) &  &  &  &  & 0.124** (0.040) \\
Glob (effFrom) &  &  &  &  & 0.002* (0.001) \\
GDP (effFrom) &  &  &  &  & 0.070*** (0.013) \\
Pop (effFrom) &  &  &  &  & -0.007ns (0.010) \\
Urb (effFrom) &  &  &  &  & -0.002*** (0.001) \\
Pubs (effFrom) &  &  &  &  & -0.003ns (0.008) \\
Deg X Dist (effFrom) &  &  &  & -0.012** (0.004) & -0.008* (0.004) \\
Deg X Lang (effFrom) &  &  &  & 0.031*** (0.009) & 0.038*** (0.009) \\
\hline
Convergence Ratio & 0.172 & 0.147 & 0.178 & 0.160 & 0.171 \\
Iterations & 6412 & 6399 & 6440 & 6472 & 6676 \\
\hline
\end{tabular}
}
\vspace{0.1cm}
\noindent\parbox{\textwidth}{\small\centering\textit{Notes: Top half are effects on the network (selection); bottom half are network effects on FWCI (influence). SEs in parentheses. ns=$p>0.05$, *=$p<0.05$, **=$p<0.01$, ***=$p<0.001$. Convergence ratio $<0.25$ is good.}}
\end{table}

In model four we add the four moderation interaction terms. On the network side we add the interaction of FWCI with geographic distance and the interaction of FWCI with common language. The distance interaction is positive and significant. As distance grows, the effect of FWCI on tie formation increases. This supports H2a and points to our main finding, that countries appear to rely on performance signals such as FWCI more heavily when potential collaborators are geographically distant. The language interaction is not significant in this model. On the performance side we add the analogous interactions of network degree with geographic remoteness and of network degree with linguistic reach. The remoteness interaction is negative and significant, and the linguistic reach interaction is positive and significant. These support H1a and H1b. The performance return to network centrality is weaker for more geographically remote countries and stronger for countries with greater linguistic reach.

Lastly, model five adds the full set of country covariates. In this model all six hypotheses are supported. FWCI performance continues to enhance tie formation and network evolution, supporting H2, and the direct effect of network centrality on performance remains positive and significant, supporting H1. The distance interaction on the network side grows in magnitude, giving stronger support for H2a. The language interaction on the network side is now negative and significant, supporting H2b. This finding emerges when publication volume is held constant. Publication volume itself attracts ties but carries no direct performance payoff, so the language moderation of partner selection is specific to the volume controls, a point we return to in the Discussion and document in the appendix robustness models. On the performance side, the two centrality interactions remain significant, supporting H1a and H1b. Among the exogenous controls, polyarchy, GDP, population size, and urbanization show positive and significant direct effects on network evolution, while globalization is significant only through its similarity effect.

\begin{table}[H]
\centering
\small
\caption{Summary of Hypothesis Tests}
\label{tab:hyp_summary}
\begin{threeparttable}

\setlength{\tabcolsep}{6pt}
\renewcommand{\arraystretch}{1.15}

\begin{adjustbox}{max width=\textwidth}
\begin{tabular}{l p{0.42\textwidth} p{0.28\textwidth} p{0.18\textwidth}}
\toprule
\textbf{Hyp.} & \textbf{Statement} & \textbf{Model Test} & \textbf{Result} \\
\midrule
H1  & \raggedright Increasing IRC network centrality enhances national S\&T performance.
& \raggedright FWCI behavior equation: \textit{IRC Net. Degree}
& \raggedright Supported ($+0.013^{***}$) \tabularnewline
H1a & \raggedright The positive effect of IRC network centrality on national S\&T research performance is weaker for geographically remote countries.
& \raggedright FWCI behavior equation: \textit{Net. Degree $\times$ Geo. Remoteness}
& \raggedright Supported ($-0.008^{*}$) \tabularnewline
H1b & \raggedright The positive effect of IRC network centrality on national S\&T research performance is stronger for countries with greater linguistic reach.
& \raggedright FWCI behavior equation: \textit{Net. Degree $\times$ Ling. Reach}
& \raggedright Supported ($+0.038^{***}$) \tabularnewline
H2  & \raggedright Increasing S\&T performance enhances IRC network evolution.
& \raggedright Network equation: \textit{FWCI (egoPlusAltX)}
& \raggedright Supported ($+0.066^{***}$) \tabularnewline
H2a & \raggedright The positive effect of performance on IRC evolution is stronger when collaborators are geographically distant.
& \raggedright Network equation: \textit{Dist $\times$ FWCI}
& \raggedright Supported ($+0.046^{***}$) \tabularnewline
H2b & \raggedright The positive effect of performance on IRC evolution is weaker when collaborators share a common language.
& \raggedright Network equation: \textit{Lang $\times$ FWCI}
& \raggedright Supported, volume-conditional ($-0.023^{***}$) \tabularnewline
\bottomrule
\end{tabular}
\end{adjustbox}

\vspace{0.2cm}
\noindent\parbox{\textwidth}{\small\centering\textit{Notes:} Estimates are Model~5 point estimates; $^{*}\,p<.05$, $^{**}\,p<.01$, $^{***}\,p<.001$. The single asterisk on H1a marks significance at the $.05$ level in Model~5 ($p=.035$), and its effect is firmer under the robustness specifications (Appendix~\ref{app:robustness}, A1--A2). ``Volume-conditional'' flags that H2b holds with publication-volume controls (Model~5) and attenuates to non-significance without them (Appendix~\ref{app:robustness}, A3).}

\end{threeparttable}
\end{table}

\section{Discussion}\label{sec:discussion}

Our results reveal the coevolutionary relationship between national research performance and the IRC network. Both the effect of network centrality on research performance (H1) and effect of research performance on network evolution (H2) are significant in our models, suggesting a dynamic feedback loop between network centrality and performance. Our results, demonstrating network effects on performance, resonate with the view that network centrality is often associated with more resources and opportunities, which in turn enhance performance \citep{powell1999network, badar_knowledge_2015, fares2021stakeholder}. Our results also align with the previous literature that scientific research is increasingly conducted by teams, as scientists working in teams are more successful in securing resources, and producing high impact research, therefore enhancing the performance of their collaborators \citep{wuchty2007increasing, Wray_2002, razzaq2022research}.

The consistent positive effect of network degree on performance across all model specifications entails important implications for science policy: countries that seek to improve their research impact may benefit from policies that actively promote IRC infrastructure, e.g. bilateral agreements, mobility programs, and cooperative funding mechanisms \citep{wagner2015continuing, davies2022research}. Such policies may have greater impact in countries that maintain relatively peripheral positions within the network \citep{ploszaj2018core}. Further, the co-evolutionary dynamic suggests that the returns for such investments may compound over time to produce non-linear performance gains \citep{merton1968matthew}.  

Beyond the direct effect of network degree, the consistent positive average alter effect suggests that collaborator performance levels affect the focal country's level of performance. Those with strong partners tend to experience stronger performance gains themselves which points to a peer influence mechanism \citep{snijders2017modeling}. Whereas degree captures the volume of a country's collaborative access, the average-alter effect shows that the \textit{quality} of partners matters independently and connection to high-performing collaborators confers even greater benefits. A policy implication is that strategic partner selection may produce greater returns \citep{liu2022scientific}. 

Results from the performance effect side of the models suggest that prior research performance attracts potential collaborators, increasing centrality in IRC network, which in turn leads to more access to collaboration opportunities and resources that further enhance performance. This self-reinforcing mechanism reflects the Matthew Effect in science \citep{merton1968matthew, azoulay2014matthew}, where initial advantages in visibility and reputation are followed by cumulative advantage over time, exacerbating disparities in global scientific network. Documenting this dynamic at the national rather than individual level connects the Matthew Effect to observations about the global stratification of science \citep{leydesdorff2008international}, implying that the international collaboration network may reinforce social structures that entrench the divide between scientifically central and peripheral nations \citep{nielsen2021global, gomez2022leading}.

Notably, the significant negative effect of performance similarity in the full model specification indicates backbone ties tend to form between countries of dissimilar performance levels rather than between peer countries. This dis-assortative pattern contrasts with the general tendency toward assortative mixing in many social networks \citep{newman2002assortative}. This finding is also consistent with a core-periphery network structure where high performers tend to connect with lower performing countries rather than forming peer cliques \citep{ploszaj2018core, borgatti2000models}. Theory suggests that peripheral nodes actively seek out connections with central nodes \citep{barabasi2002evolution}, while central nodes reciprocate to gain access to novel resources at the edges of the network \citep{wagner2005network}. This dynamic reinforces the co-evolutionary logic as performance-driven tie formation across the core-periphery divide may simultaneously enhance the performance of peripheral partners through knowledge transfers and resource access. This finding also complements the broader social effect of performance by pointing to an additional bridging effect across performance gaps, i.e. brokerage across structural holes \citep{burt2004structural, wang2023effect}. Countries that link otherwise weakly connected regions may gain access to non-redundant information and resources, which complements the structural core-periphery account.  

Geographic distance has a consistent and strongly negative effect on network evolution across every specification. This cuts against the view that online databases and virtual communication have dissolved the effect of geography on science \citep{gui2019globalization, yan2011institutional}. If that were the case, then the distance effect would be weak. Instead it is large, negative, and significant in every model. Thus, the results support the conclusion that geography still structures who collaborates with whom \citep{katz1994geographical, hoekman2010research}, even after three decades of expanding digital infrastructure.

The four moderation tests connect our results to several established literatures. On the returns side, the performance return to network centrality declines with geographic remoteness and increases with linguistic reach. The decline with remoteness fits the coordination-cost account of distributed collaboration, where geographic separation raises the cost of joint work and erodes its payoff \citep{cummings2007coordination, olson2000distance}. The increase with linguistic reach points the same way from the communication side. A shared language lowers the cost of exchanging tacit and nuanced knowledge between collaborators \citep{hwang2013effects, tenzer2013leading}. Linguistic reach and network access are therefore complementary. Centrality increases opportunities to reach new collaborators, but converting opportunity into performance depends on knowledge flow across the tie. Thus, a country's central position converts into larger performance gains when linguistic proximity increases.

On the selection side, the result showed that where collaborations do form across distance, the effect of national performance on tie formation is stronger. Thus, distance is not only a barrier to collaboration. It also appears to be a condition under which observable performance signals become stronger. We theorize that magnified information asymmetries at the point of partner selection may increase the salience of observable performance signals \citep{spence1978job, connelly2011signaling, podolny2001networks}. In other words, when scientists lack direct information on their distant collaborators, they may rely more heavily on proxies for performance such as citations \citep{abramo2024moderating, onuchic2023signaling}. However, this interpretation is retrospective, leaving room for alternative mechanisms and highlighting the need for future research to causally test whether performance signaling actually mediates the effect of distance. 

The language interaction on the selection side (H2b) is more complex. It is null until publication volume enters the model and becomes significant and negative afterwards. Otherwise identical specifications without the volume term show no such effect (Appendix Model~A3). It thus appears that publication volume suppresses language moderation. Prolific countries are highly visible within their own language communities \citep{amano2016languages, hwang2005inferior}, and only once volume is held constant does the signal-substituting role of shared language emerge. We therefore treat support as conditional on the volume controls.

Taken together, our moderation results add to the proximity literature, which largely treats proximity as a force acting on collaboration \citep{boschma2005proximity, frenken2009spatial}. Our results instead point to a broader set of effects across the collaboration process. Information asymmetry appears to govern partner \textit{selection}, where performance signals matter most when direct information about distant partners is scarce. Communication and coordination appear to govern the \textit{returns} to collaboration, which a shared language amplifies and distance attenuates.

Lastly, our research introduces a novel application of the \texttt{RSiena} SAOM framework to test reciprocal relationship between research performance and network evolution. While existing studies focus on testing either the effect of network position on performance or productivity on network position, SAOM enables testing the co-evolution of both network and performance in a robust way. In traditional regression models, reverse causality often leads to endogeneity problems since prior performance influences network evolution and network centrality also affects subsequent research performance at the same time. SAOM addresses such endogeneity issue by explicitly modeling co-evolution of network centrality and countries' collaborative behavior over time. The model thus captures co-evolutionary dynamics, reducing bias caused by endogeneity.

\subsection{Limitations \& Future Directions}

Several limitations qualify these results and point to future work. The study is observational, so the estimates identify patterns of co-evolution rather than experimentally isolated causes. The co-evolution model reduces the endogeneity bias that arises when performance and network position are estimated in separate equations, but it does not by itself establish a causal mechanism. This matters most for the moderation results. The design identifies the moderation pattern, yet several processes could generate it. Performance signaling under information asymmetry is one explanation, but bilateral agreements, complementary national research portfolios, and unequal access to funding may be observationally equivalent at this level of aggregation. Discriminating among these accounts with future tests is needed to establish which mechanism operates.

As mentioned earlier, the SAOM imposes two data requirements that shape the analysis. Network ties must be binary, so the valued co-authorship matrices were reduced to backbones with a disparity filter \citep{serrano2009extracting, neal2014backbone}. Performance must be discrete, so fractional FWCI was discretized into ordinal levels, since continuous specifications of the behavior equation did not converge \citep{niezink2019discrete}. Both steps discard information. We assessed sensitivity to the binarization threshold by re-estimating the full model on looser backbones, and every hypothesis retained its sign and significance (Appendix~\ref{app:robustness}, Model~A2).

\section{Conclusion}

Until this point, scholars of science and technology policy have analyzed the relationship between international research collaboration (IRC) and research performance uni-directionally. This paper leverages the tools of network science to test reciprocal pathways in a single model and explicitly models the endogeneity between collaboration and performance. In a thirty year network panel of 172 countries, we demonstrate with a single unified model that collaboration enhances research performance \textit{and} that research performance enhances collaboration. In other words, international collaboration and national performance co-evolve dynamically across time and space.

This co-evolution is shaped by geography and language in important ways. Distance and shared language appear to condition collaborator selection as well as the influence of collaboration on citation performance. However, these effects operate at different stages. When direct information about distant partners is scarce, observable performance acts as a proxy signal for partner quality, so distance strengthens the effect of performance on partner selection. Once collaborations form, a shared language raises the returns to network access while distance lowers them. This difference between the two stages is part of our central theoretical contribution.

These dynamics also carry implications for science policy. The co-evolutionary cycle suggests that investments in international research collaboration attract further collaborative opportunities, which in turn generate additional performance benefits. Early and sustained investment may therefore be particularly consequential for countries seeking to improve their position in the global network of science, and investments that enhance network positioning in the network may matter more than simply increasing productivity. However, because advantage tends to accumulate, the same endogenous cycles may further entrench the divide between scientifically central and peripheral nations. 

\appendix
\renewcommand{\thetable}{A\arabic{table}}
\renewcommand{\thefigure}{A\arabic{figure}}
\setcounter{table}{0}
\setcounter{figure}{0}
\section{Appendix}
\label{app:robustness}

Table~\ref{tab:robustness} reports three robustness variants of the full model (Model~5): Model~A1 substitutes R\&D intensity for GDP per capita; Model~A2 re-estimates on the looser $\alpha = 0.10$ disparity backbones; and Model~A3 removes the publication volume terms to isolate their effect on H2b. Every hypothesis effect retains its predicted sign and significance across all three variants. RDGDP has a large number of zeros for non-reporting countries. These are recoded to zero for this indicator, following \citet{wagner2024science}, on the assumption that non-reporting signals a negligible formal R\&D system. Taiwan is absent from the World Bank by non-membership and is retained as missing.

Further testing indicated that including publication volume degrades the model's goodness of fit \citep{lospinoso2019goodness}. Simulating networks and behavior at the estimated parameters, Model~5 departs from the observed degree distribution by generating more isolated countries than are observed, and from the observed triad census by under-producing complete triads ($p < .001$). Model~A3 is identical apart from the volume terms and reproduces the triad census adequately ($p = .204$), which localizes that misfit to those terms. These terms are retained in Model~5 because they separate the performance effect from sheer publication output.

\begin{table}[H]
\centering
\small
\caption{Robustness Models (variants of Model~5; compare Model~5 in Table~\ref{tab:Co-Evolution Models})}
\label{tab:robustness}
\adjustbox{max width=\textwidth}{%
\begin{tabular}{lccc}
Variable & A1 (R\&D int.) & A2 ($\alpha{=}.10$) & A3 (no pubs) \\
\hline
\multicolumn{4}{c}{Effects on Network Dynamics (Selection)} \\
\hline
Intercept & -3.369*** (0.062) & -3.034*** (0.056) & -3.096*** (0.054) \\
Transitivity (gwesp) & 0.405*** (0.035) & 0.379*** (0.031) & 0.353*** (0.031) \\
Pref. Attach (degPlus) & 0.009*** (0.000) & 0.009*** (0.000) & 0.012*** (0.000) \\
Dist (X) & -0.996*** (0.024) & -0.949*** (0.018) & -0.962*** (0.023) \\
Lang (X) & 1.060*** (0.030) & 0.967*** (0.025) & 0.876*** (0.028) \\
FWCI (egoPlusAltX) & 0.072*** (0.007) & 0.057*** (0.005) & 0.067*** (0.006) \\
FWCI (simX) & -0.172* (0.067) & -0.191*** (0.053) & -0.240*** (0.067) \\
Poly (egoPlusAltX) & 0.333*** (0.036) & 0.399*** (0.029) & 0.416*** (0.034) \\
Poly (simX) & 0.295*** (0.048) & 0.243*** (0.038) & 0.264*** (0.049) \\
Glob (egoPlusAltX) & 0.001ns (0.001) & -0.000ns (0.001) & -0.000ns (0.001) \\
Glob (simX) & 0.554*** (0.066) & 0.532*** (0.058) & 0.796*** (0.071) \\
GDP (egoPlusAltX) &  & 0.053*** (0.012) & 0.136*** (0.013) \\
GDP (simX) &  & 0.209* (0.086) & -0.483*** (0.105) \\
R\&D int. (egoPlusAltX) & -0.014ns (0.015) &  &  \\
R\&D int. (simX) & 0.680*** (0.104) &  &  \\
Pop (egoPlusAltX) & 0.166*** (0.010) & 0.174*** (0.009) & 0.285*** (0.007) \\
Pop (simX) & 0.629*** (0.082) & 0.509*** (0.064) & -0.209** (0.072) \\
Urb (egoPlusAltX) & 0.003*** (0.000) & 0.000ns (0.000) & 0.002*** (0.001) \\
Urb (simX) & -0.030ns (0.064) & 0.058ns (0.058) & -0.018ns (0.071) \\
Pubs (egoPlusAltX) & 0.162*** (0.009) & 0.115*** (0.007) &  \\
Pubs (simX) & -1.940*** (0.101) & -1.248*** (0.076) &  \\
Dist x FWCI (egoPlusAltX) & 0.045*** (0.005) & 0.042*** (0.004) & 0.044*** (0.005) \\
Lang x FWCI (egoPlusAltX) & -0.030*** (0.007) & -0.018** (0.006) & 0.008ns (0.007) \\
\hline
\multicolumn{4}{c}{Effects on Behavior Dynamics (Influence)} \\
\hline
Linear shape & -0.262*** (0.032) & -0.282*** (0.036) & -0.263*** (0.032) \\
Quad. shape & -0.014*** (0.002) & -0.013*** (0.002) & -0.013*** (0.002) \\
IRC Net. Deg. (outdeg) & 0.012*** (0.001) & 0.010*** (0.001) & 0.013*** (0.001) \\
IRC Avg. Alter (avAlt) & 0.078*** (0.010) & 0.081*** (0.012) & 0.075*** (0.010) \\
Dist (effFrom) & 0.163*** (0.044) & 0.121** (0.045) & 0.104* (0.043) \\
Lang (effFrom) & 0.067ns (0.098) & 0.049ns (0.101) & 0.081ns (0.099) \\
Poly (effFrom) & 0.086* (0.040) & 0.138*** (0.040) & 0.120** (0.038) \\
Glob (effFrom) & 0.003** (0.001) & 0.002* (0.001) & 0.002ns (0.001) \\
GDP (effFrom) &  & 0.068*** (0.013) & 0.067*** (0.012) \\
R\&D int. (effFrom) & 0.130*** (0.019) &  &  \\
Pop (effFrom) & -0.026** (0.009) & -0.003ns (0.010) & -0.011ns (0.007) \\
Urb (effFrom) & -0.000ns (0.001) & -0.002*** (0.001) & -0.002*** (0.001) \\
Pubs (effFrom) & 0.005ns (0.007) & -0.003ns (0.008) &  \\
Deg X Dist (effFrom) & -0.011** (0.004) & -0.007** (0.003) & -0.008* (0.004) \\
Deg X Lang (effFrom) & 0.047*** (0.009) & 0.032*** (0.006) & 0.037*** (0.009) \\
\hline
Convergence Ratio & 0.169 & 0.182 & 0.151 \\
\hline
\end{tabular}}
\begin{center}\small{\textit{Notes:} SEs in parentheses. ns\,$p>.05$, *\,$p<.05$, **\,$p<.01$, ***\,$p<.001$. All models are variants of Model~5 (Table~\ref{tab:Co-Evolution Models}): A1 substitutes R\&D intensity for GDP per capita; A2 re-estimates on the looser $\alpha=0.10$ disparity backbones; A3 removes the publication volume terms.}\end{center}
\end{table}


\begin{thebibliography}{}

\bibitem [\protect \citeauthoryear {%
Abramo%
, Apponi%
\BCBL {}\ \BBA {} D'Angelo%
}{%
Abramo%
\ \protect \BOthers {.}}{%
{\protect \APACyear {2024}}%
}]{%
abramo2024moderating}
\APACinsertmetastar {%
abramo2024moderating}%
\begin{APACrefauthors}%
Abramo, G.%
, Apponi, F.%
\BCBL {}\ \BBA {} D'Angelo, C\BPBI A.%
\end{APACrefauthors}%
\unskip\
\newblock
\APACrefYearMonthDay{2024}{}{}.
\newblock
{\BBOQ}\APACrefatitle {The moderating role of the territorial research infrastructure on the geographic proximity effect in research collaborations: a regional-based view} {The moderating role of the territorial research infrastructure on the geographic proximity effect in research collaborations: a regional-based view}.{\BBCQ}
\newblock
\APACjournalVolNumPages{Scientometrics}{129}{6}{3149--3168}.
\PrintBackRefs{\CurrentBib}

\bibitem [\protect \citeauthoryear {%
Abramo%
, D'Angelo%
\BCBL {}\ \BBA {} Di~Costa%
}{%
Abramo%
\ \protect \BOthers {.}}{%
{\protect \APACyear {2009}}%
}]{%
abramo2009research}
\APACinsertmetastar {%
abramo2009research}%
\begin{APACrefauthors}%
Abramo, G.%
, D'Angelo, C\BPBI A.%
\BCBL {}\ \BBA {} Di~Costa, F.%
\end{APACrefauthors}%
\unskip\
\newblock
\APACrefYearMonthDay{2009}{}{}.
\newblock
{\BBOQ}\APACrefatitle {Research collaboration and productivity: is there correlation?} {Research collaboration and productivity: is there correlation?}{\BBCQ}
\newblock
\APACjournalVolNumPages{Higher education}{57}{}{155--171}.
\PrintBackRefs{\CurrentBib}

\bibitem [\protect \citeauthoryear {%
Abramo%
, D'Angelo%
\BCBL {}\ \BBA {} Di~Costa%
}{%
Abramo%
\ \protect \BOthers {.}}{%
{\protect \APACyear {2019}}%
}]{%
abramo2019collaboration}
\APACinsertmetastar {%
abramo2019collaboration}%
\begin{APACrefauthors}%
Abramo, G.%
, D'Angelo, C\BPBI A.%
\BCBL {}\ \BBA {} Di~Costa, F.%
\end{APACrefauthors}%
\unskip\
\newblock
\APACrefYearMonthDay{2019}{}{}.
\newblock
{\BBOQ}\APACrefatitle {The collaboration behavior of top scientists} {The collaboration behavior of top scientists}.{\BBCQ}
\newblock
\APACjournalVolNumPages{Scientometrics}{118}{1}{215--232}.
\PrintBackRefs{\CurrentBib}

\bibitem [\protect \citeauthoryear {%
Abramo%
, D'Angelo%
\BCBL {}\ \BBA {} Di~Costa%
}{%
Abramo%
\ \protect \BOthers {.}}{%
{\protect \APACyear {2020}}%
{\protect \APACexlab {{\protect \BCnt {1}}}}}]{%
abramo2020knowledge}
\APACinsertmetastar {%
abramo2020knowledge}%
\begin{APACrefauthors}%
Abramo, G.%
, D'Angelo, C\BPBI A.%
\BCBL {}\ \BBA {} Di~Costa, F.%
\end{APACrefauthors}%
\unskip\
\newblock
\APACrefYearMonthDay{2020{\protect \BCnt {1}}}{}{}.
\newblock
{\BBOQ}\APACrefatitle {Knowledge spillovers: Does the geographic proximity effect decay over time? A discipline-level analysis, accounting for cognitive proximity, with and without self-citations} {Knowledge spillovers: Does the geographic proximity effect decay over time? a discipline-level analysis, accounting for cognitive proximity, with and without self-citations}.{\BBCQ}
\newblock
\APACjournalVolNumPages{Journal of Informetrics}{14}{4}{101072}.
\PrintBackRefs{\CurrentBib}

\bibitem [\protect \citeauthoryear {%
Abramo%
, D'Angelo%
\BCBL {}\ \BBA {} Di~Costa%
}{%
Abramo%
\ \protect \BOthers {.}}{%
{\protect \APACyear {2020}}%
{\protect \APACexlab {{\protect \BCnt {2}}}}}]{%
abramo2020role}
\APACinsertmetastar {%
abramo2020role}%
\begin{APACrefauthors}%
Abramo, G.%
, D'Angelo, C\BPBI A.%
\BCBL {}\ \BBA {} Di~Costa, F.%
\end{APACrefauthors}%
\unskip\
\newblock
\APACrefYearMonthDay{2020{\protect \BCnt {2}}}{}{}.
\newblock
{\BBOQ}\APACrefatitle {The role of geographical proximity in knowledge diffusion, measured by citations to scientific literature} {The role of geographical proximity in knowledge diffusion, measured by citations to scientific literature}.{\BBCQ}
\newblock
\APACjournalVolNumPages{Journal of Informetrics}{14}{1}{101010}.
\PrintBackRefs{\CurrentBib}

\bibitem [\protect \citeauthoryear {%
Abramo%
, D'Angelo%
\BCBL {}\ \BBA {} Solazzi%
}{%
Abramo%
\ \protect \BOthers {.}}{%
{\protect \APACyear {2011}}%
}]{%
abramo2011relationship}
\APACinsertmetastar {%
abramo2011relationship}%
\begin{APACrefauthors}%
Abramo, G.%
, D'Angelo, C\BPBI A.%
\BCBL {}\ \BBA {} Solazzi, M.%
\end{APACrefauthors}%
\unskip\
\newblock
\APACrefYearMonthDay{2011}{}{}.
\newblock
{\BBOQ}\APACrefatitle {The relationship between scientists' research performance and the degree of internationalization of their research} {The relationship between scientists' research performance and the degree of internationalization of their research}.{\BBCQ}
\newblock
\APACjournalVolNumPages{Scientometrics}{86}{3}{629--643}.
\PrintBackRefs{\CurrentBib}

\bibitem [\protect \citeauthoryear {%
Adams%
}{%
Adams%
}{%
{\protect \APACyear {2013}}%
}]{%
adams2013fourth}
\APACinsertmetastar {%
adams2013fourth}%
\begin{APACrefauthors}%
Adams, J.%
\end{APACrefauthors}%
\unskip\
\newblock
\APACrefYearMonthDay{2013}{}{}.
\newblock
{\BBOQ}\APACrefatitle {The fourth age of research} {The fourth age of research}.{\BBCQ}
\newblock
\APACjournalVolNumPages{Nature}{497}{7451}{557--560}.
\PrintBackRefs{\CurrentBib}

\bibitem [\protect \citeauthoryear {%
Adams%
\ \BBA {} Szomszor%
}{%
Adams%
\ \BBA {} Szomszor%
}{%
{\protect \APACyear {2024}}%
}]{%
adams2024national}
\APACinsertmetastar {%
adams2024national}%
\begin{APACrefauthors}%
Adams, J.%
\BCBT {}\ \BBA {} Szomszor, M.%
\end{APACrefauthors}%
\unskip\
\newblock
\APACrefYearMonthDay{2024}{}{}.
\newblock
{\BBOQ}\APACrefatitle {National research impact is driven by global collaboration, not rising performance} {National research impact is driven by global collaboration, not rising performance}.{\BBCQ}
\newblock
\APACjournalVolNumPages{Scientometrics}{129}{5}{2883--2896}.
\PrintBackRefs{\CurrentBib}

\bibitem [\protect \citeauthoryear {%
Aksnes%
, Langfeldt%
\BCBL {}\ \BBA {} Wouters%
}{%
Aksnes%
\ \protect \BOthers {.}}{%
{\protect \APACyear {2019}}%
}]{%
aksnes2019citations}
\APACinsertmetastar {%
aksnes2019citations}%
\begin{APACrefauthors}%
Aksnes, D\BPBI W.%
, Langfeldt, L.%
\BCBL {}\ \BBA {} Wouters, P.%
\end{APACrefauthors}%
\unskip\
\newblock
\APACrefYearMonthDay{2019}{}{}.
\newblock
{\BBOQ}\APACrefatitle {Citations, citation indicators, and research quality: An overview of basic concepts and theories} {Citations, citation indicators, and research quality: An overview of basic concepts and theories}.{\BBCQ}
\newblock
\APACjournalVolNumPages{Sage Open}{9}{1}{2158244019829575}.
\PrintBackRefs{\CurrentBib}

\bibitem [\protect \citeauthoryear {%
Amano%
, Gonz{\'a}lez-Varo%
\BCBL {}\ \BBA {} Sutherland%
}{%
Amano%
\ \protect \BOthers {.}}{%
{\protect \APACyear {2016}}%
}]{%
amano2016languages}
\APACinsertmetastar {%
amano2016languages}%
\begin{APACrefauthors}%
Amano, T.%
, Gonz{\'a}lez-Varo, J\BPBI P.%
\BCBL {}\ \BBA {} Sutherland, W\BPBI J.%
\end{APACrefauthors}%
\unskip\
\newblock
\APACrefYearMonthDay{2016}{}{}.
\newblock
{\BBOQ}\APACrefatitle {Languages are still a major barrier to global science} {Languages are still a major barrier to global science}.{\BBCQ}
\newblock
\APACjournalVolNumPages{PLoS Biology}{14}{12}{e2000933}.
\PrintBackRefs{\CurrentBib}

\bibitem [\protect \citeauthoryear {%
Auranen%
\ \BBA {} Nieminen%
}{%
Auranen%
\ \BBA {} Nieminen%
}{%
{\protect \APACyear {2010}}%
}]{%
auranen2010university}
\APACinsertmetastar {%
auranen2010university}%
\begin{APACrefauthors}%
Auranen, O.%
\BCBT {}\ \BBA {} Nieminen, M.%
\end{APACrefauthors}%
\unskip\
\newblock
\APACrefYearMonthDay{2010}{}{}.
\newblock
{\BBOQ}\APACrefatitle {University research funding and publication performance---An international comparison} {University research funding and publication performance---an international comparison}.{\BBCQ}
\newblock
\APACjournalVolNumPages{Research Policy}{39}{6}{822--834}.
\PrintBackRefs{\CurrentBib}

\bibitem [\protect \citeauthoryear {%
Azoulay%
, Stuart%
\BCBL {}\ \BBA {} Wang%
}{%
Azoulay%
\ \protect \BOthers {.}}{%
{\protect \APACyear {2014}}%
}]{%
azoulay2014matthew}
\APACinsertmetastar {%
azoulay2014matthew}%
\begin{APACrefauthors}%
Azoulay, P.%
, Stuart, T.%
\BCBL {}\ \BBA {} Wang, Y.%
\end{APACrefauthors}%
\unskip\
\newblock
\APACrefYearMonthDay{2014}{}{}.
\newblock
{\BBOQ}\APACrefatitle {Matthew: Effect or fable?} {Matthew: Effect or fable?}{\BBCQ}
\newblock
\APACjournalVolNumPages{Management Science}{60}{1}{92--109}.
\PrintBackRefs{\CurrentBib}

\bibitem [\protect \citeauthoryear {%
Badar%
, Hite%
\BCBL {}\ \BBA {} Ashraf%
}{%
Badar%
\ \protect \BOthers {.}}{%
{\protect \APACyear {2015}}%
}]{%
badar_knowledge_2015}
\APACinsertmetastar {%
badar_knowledge_2015}%
\begin{APACrefauthors}%
Badar, K.%
, Hite, J\BPBI M.%
\BCBL {}\ \BBA {} Ashraf, N.%
\end{APACrefauthors}%
\unskip\
\newblock
\APACrefYearMonthDay{2015}{}{}.
\newblock
{\BBOQ}\APACrefatitle {Knowledge network centrality, formal rank and research performance: evidence for curvilinear and interaction effects} {Knowledge network centrality, formal rank and research performance: evidence for curvilinear and interaction effects}.{\BBCQ}
\newblock
\APACjournalVolNumPages{Scientometrics}{105}{3}{1553--1576}.
\newblock
\begin{APACrefDOI} \doi{10.1007/s11192-015-1652-0} \end{APACrefDOI}
\PrintBackRefs{\CurrentBib}

\bibitem [\protect \citeauthoryear {%
Barab{\^a}si%
\ \protect \BOthers {.}}{%
Barab{\^a}si%
\ \protect \BOthers {.}}{%
{\protect \APACyear {2002}}%
}]{%
barabasi2002evolution}
\APACinsertmetastar {%
barabasi2002evolution}%
\begin{APACrefauthors}%
Barab{\^a}si, A\BHBI L.%
, Jeong, H.%
, N{\'e}da, Z.%
, Ravasz, E.%
, Schubert, A.%
\BCBL {}\ \BBA {} Vicsek, T.%
\end{APACrefauthors}%
\unskip\
\newblock
\APACrefYearMonthDay{2002}{}{}.
\newblock
{\BBOQ}\APACrefatitle {Evolution of the social network of scientific collaborations} {Evolution of the social network of scientific collaborations}.{\BBCQ}
\newblock
\APACjournalVolNumPages{Physica A: Statistical Mechanics and its Applications}{311}{3-4}{590--614}.
\PrintBackRefs{\CurrentBib}

\bibitem [\protect \citeauthoryear {%
Beaver%
\ \BBA {} Rosen%
}{%
Beaver%
\ \BBA {} Rosen%
}{%
{\protect \APACyear {1978}}%
}]{%
beaver1978studies}
\APACinsertmetastar {%
beaver1978studies}%
\begin{APACrefauthors}%
Beaver, D.%
\BCBT {}\ \BBA {} Rosen, R.%
\end{APACrefauthors}%
\unskip\
\newblock
\APACrefYearMonthDay{1978}{}{}.
\newblock
{\BBOQ}\APACrefatitle {Studies in scientific collaboration: Part {I}. The professional origins of scientific co-authorship} {Studies in scientific collaboration: Part {I}. the professional origins of scientific co-authorship}.{\BBCQ}
\newblock
\APACjournalVolNumPages{Scientometrics}{1}{1}{65--84}.
\PrintBackRefs{\CurrentBib}

\bibitem [\protect \citeauthoryear {%
Beaver%
\ \BBA {} Rosen%
}{%
Beaver%
\ \BBA {} Rosen%
}{%
{\protect \APACyear {1979}}%
}]{%
beaver1979studies}
\APACinsertmetastar {%
beaver1979studies}%
\begin{APACrefauthors}%
Beaver, D.%
\BCBT {}\ \BBA {} Rosen, R.%
\end{APACrefauthors}%
\unskip\
\newblock
\APACrefYearMonthDay{1979}{}{}.
\newblock
{\BBOQ}\APACrefatitle {Studies in scientific collaboration: Part {II}. Scientific co-authorship, research productivity and visibility in the {French} scientific elite, 1799--1830} {Studies in scientific collaboration: Part {II}. scientific co-authorship, research productivity and visibility in the {French} scientific elite, 1799--1830}.{\BBCQ}
\newblock
\APACjournalVolNumPages{Scientometrics}{1}{2}{133--149}.
\PrintBackRefs{\CurrentBib}

\bibitem [\protect \citeauthoryear {%
Borgatti%
\ \BBA {} Everett%
}{%
Borgatti%
\ \BBA {} Everett%
}{%
{\protect \APACyear {2000}}%
}]{%
borgatti2000models}
\APACinsertmetastar {%
borgatti2000models}%
\begin{APACrefauthors}%
Borgatti, S\BPBI P.%
\BCBT {}\ \BBA {} Everett, M\BPBI G.%
\end{APACrefauthors}%
\unskip\
\newblock
\APACrefYearMonthDay{2000}{}{}.
\newblock
{\BBOQ}\APACrefatitle {Models of core/periphery structures} {Models of core/periphery structures}.{\BBCQ}
\newblock
\APACjournalVolNumPages{Social Networks}{21}{4}{375--395}.
\PrintBackRefs{\CurrentBib}

\bibitem [\protect \citeauthoryear {%
Boschma%
}{%
Boschma%
}{%
{\protect \APACyear {2005}}%
}]{%
boschma2005proximity}
\APACinsertmetastar {%
boschma2005proximity}%
\begin{APACrefauthors}%
Boschma, R\BPBI A.%
\end{APACrefauthors}%
\unskip\
\newblock
\APACrefYearMonthDay{2005}{}{}.
\newblock
{\BBOQ}\APACrefatitle {Proximity and innovation: A critical assessment} {Proximity and innovation: A critical assessment}.{\BBCQ}
\newblock
\APACjournalVolNumPages{Regional Studies}{39}{1}{61--74}.
\PrintBackRefs{\CurrentBib}

\bibitem [\protect \citeauthoryear {%
Burt%
}{%
Burt%
}{%
{\protect \APACyear {2004}}%
}]{%
burt2004structural}
\APACinsertmetastar {%
burt2004structural}%
\begin{APACrefauthors}%
Burt, R\BPBI S.%
\end{APACrefauthors}%
\unskip\
\newblock
\APACrefYearMonthDay{2004}{}{}.
\newblock
{\BBOQ}\APACrefatitle {Structural holes and good ideas} {Structural holes and good ideas}.{\BBCQ}
\newblock
\APACjournalVolNumPages{American Journal of Sociology}{110}{2}{349--399}.
\PrintBackRefs{\CurrentBib}

\bibitem [\protect \citeauthoryear {%
Chen%
, Ding%
, Zhao%
, Guo%
\BCBL {}\ \BBA {} Ning%
}{%
Chen%
\ \protect \BOthers {.}}{%
{\protect \APACyear {2025}}%
}]{%
chen2025benefits}
\APACinsertmetastar {%
chen2025benefits}%
\begin{APACrefauthors}%
Chen, K.%
, Ding, Y.%
, Zhao, B.%
, Guo, R.%
\BCBL {}\ \BBA {} Ning, L.%
\end{APACrefauthors}%
\unskip\
\newblock
\APACrefYearMonthDay{2025}{}{}.
\newblock
{\BBOQ}\APACrefatitle {Benefits beyond the local network: Does indirect international collaboration ties contribute to research performance for young scientists?} {Benefits beyond the local network: Does indirect international collaboration ties contribute to research performance for young scientists?}{\BBCQ}
\newblock
\APACjournalVolNumPages{Research Policy}{54}{5}{105233}.
\PrintBackRefs{\CurrentBib}

\bibitem [\protect \citeauthoryear {%
Chen%
, Zhang%
\BCBL {}\ \BBA {} Fu%
}{%
Chen%
\ \protect \BOthers {.}}{%
{\protect \APACyear {2019}}%
}]{%
chen2019international}
\APACinsertmetastar {%
chen2019international}%
\begin{APACrefauthors}%
Chen, K.%
, Zhang, Y.%
\BCBL {}\ \BBA {} Fu, X.%
\end{APACrefauthors}%
\unskip\
\newblock
\APACrefYearMonthDay{2019}{}{}.
\newblock
{\BBOQ}\APACrefatitle {International research collaboration: An emerging domain of innovation studies?} {International research collaboration: An emerging domain of innovation studies?}{\BBCQ}
\newblock
\APACjournalVolNumPages{Research Policy}{48}{1}{149--168}.
\PrintBackRefs{\CurrentBib}

\bibitem [\protect \citeauthoryear {%
Coleman%
}{%
Coleman%
}{%
{\protect \APACyear {1990}}%
}]{%
coleman1990foundations}
\APACinsertmetastar {%
coleman1990foundations}%
\begin{APACrefauthors}%
Coleman, J\BPBI S.%
\end{APACrefauthors}%
\unskip\
\newblock
\APACrefYear{1990}.
\newblock
\APACrefbtitle {Foundations of Social Theory} {Foundations of social theory}.
\newblock
\APACaddressPublisher{}{Harvard University Press}.
\PrintBackRefs{\CurrentBib}

\bibitem [\protect \citeauthoryear {%
Connelly%
, Certo%
, Ireland%
\BCBL {}\ \BBA {} Reutzel%
}{%
Connelly%
\ \protect \BOthers {.}}{%
{\protect \APACyear {2011}}%
}]{%
connelly2011signaling}
\APACinsertmetastar {%
connelly2011signaling}%
\begin{APACrefauthors}%
Connelly, B\BPBI L.%
, Certo, S\BPBI T.%
, Ireland, R\BPBI D.%
\BCBL {}\ \BBA {} Reutzel, C\BPBI R.%
\end{APACrefauthors}%
\unskip\
\newblock
\APACrefYearMonthDay{2011}{}{}.
\newblock
{\BBOQ}\APACrefatitle {Signaling theory: A review and assessment} {Signaling theory: A review and assessment}.{\BBCQ}
\newblock
\APACjournalVolNumPages{Journal of Management}{37}{1}{39--67}.
\PrintBackRefs{\CurrentBib}

\bibitem [\protect \citeauthoryear {%
Coppedge%
}{%
Coppedge%
}{%
{\protect \APACyear {2024}}%
}]{%
VDEM2024}
\APACinsertmetastar {%
VDEM2024}%
\begin{APACrefauthors}%
Coppedge, M.%
\end{APACrefauthors}%
\unskip\
\newblock
\APACrefYearMonthDay{2024}{}{}.
\newblock
\APACrefbtitle {{Varieties of Democracy Codebook v14}.} {{Varieties of Democracy Codebook v14}.}
\newblock
\APAChowpublished {\url{https://v-dem.net/documents/38/V-Dem_Codebook_v14.pdf}}.
\PrintBackRefs{\CurrentBib}

\bibitem [\protect \citeauthoryear {%
Cranmer%
, Desmarais%
\BCBL {}\ \BBA {} Morgan%
}{%
Cranmer%
\ \protect \BOthers {.}}{%
{\protect \APACyear {2020}}%
}]{%
cranmer2020inferential}
\APACinsertmetastar {%
cranmer2020inferential}%
\begin{APACrefauthors}%
Cranmer, S\BPBI J.%
, Desmarais, B\BPBI A.%
\BCBL {}\ \BBA {} Morgan, J\BPBI W.%
\end{APACrefauthors}%
\unskip\
\newblock
\APACrefYear{2020}.
\newblock
\APACrefbtitle {Inferential Network Analysis} {Inferential network analysis}.
\newblock
\APACaddressPublisher{}{Cambridge University Press}.
\PrintBackRefs{\CurrentBib}

\bibitem [\protect \citeauthoryear {%
Cummings%
\ \BBA {} Kiesler%
}{%
Cummings%
\ \BBA {} Kiesler%
}{%
{\protect \APACyear {2007}}%
}]{%
cummings2007coordination}
\APACinsertmetastar {%
cummings2007coordination}%
\begin{APACrefauthors}%
Cummings, J\BPBI N.%
\BCBT {}\ \BBA {} Kiesler, S.%
\end{APACrefauthors}%
\unskip\
\newblock
\APACrefYearMonthDay{2007}{}{}.
\newblock
{\BBOQ}\APACrefatitle {Coordination costs and project outcomes in multi-university collaborations} {Coordination costs and project outcomes in multi-university collaborations}.{\BBCQ}
\newblock
\APACjournalVolNumPages{Research Policy}{36}{10}{1620--1634}.
\PrintBackRefs{\CurrentBib}

\bibitem [\protect \citeauthoryear {%
Davies%
, Gush%
, Hendy%
\BCBL {}\ \BBA {} Jaffe%
}{%
Davies%
\ \protect \BOthers {.}}{%
{\protect \APACyear {2022}}%
}]{%
davies2022research}
\APACinsertmetastar {%
davies2022research}%
\begin{APACrefauthors}%
Davies, B.%
, Gush, J.%
, Hendy, S\BPBI C.%
\BCBL {}\ \BBA {} Jaffe, A\BPBI B.%
\end{APACrefauthors}%
\unskip\
\newblock
\APACrefYearMonthDay{2022}{}{}.
\newblock
{\BBOQ}\APACrefatitle {Research funding and collaboration} {Research funding and collaboration}.{\BBCQ}
\newblock
\APACjournalVolNumPages{Research Policy}{51}{2}{104421}.
\PrintBackRefs{\CurrentBib}

\bibitem [\protect \citeauthoryear {%
de Frutos-Beliz{\'o}n%
, Garc{\'\i}a-Carbonell%
, Guerrero-Alba%
\BCBL {}\ \BBA {} S{\'a}nchez-Gardey%
}{%
de Frutos-Beliz{\'o}n%
\ \protect \BOthers {.}}{%
{\protect \APACyear {2024}}%
}]{%
de2024empirical}
\APACinsertmetastar {%
de2024empirical}%
\begin{APACrefauthors}%
de Frutos-Beliz{\'o}n, J.%
, Garc{\'\i}a-Carbonell, N.%
, Guerrero-Alba, F.%
\BCBL {}\ \BBA {} S{\'a}nchez-Gardey, G.%
\end{APACrefauthors}%
\unskip\
\newblock
\APACrefYearMonthDay{2024}{}{}.
\newblock
{\BBOQ}\APACrefatitle {An empirical analysis of individual and collective determinants of international research collaboration} {An empirical analysis of individual and collective determinants of international research collaboration}.{\BBCQ}
\newblock
\APACjournalVolNumPages{Scientometrics}{129}{5}{2749--2770}.
\PrintBackRefs{\CurrentBib}

\bibitem [\protect \citeauthoryear {%
Dreher%
}{%
Dreher%
}{%
{\protect \APACyear {2006}}%
}]{%
dreher2006does}
\APACinsertmetastar {%
dreher2006does}%
\begin{APACrefauthors}%
Dreher, A.%
\end{APACrefauthors}%
\unskip\
\newblock
\APACrefYearMonthDay{2006}{}{}.
\newblock
{\BBOQ}\APACrefatitle {Does globalization affect growth? {Evidence} from a new index of globalization} {Does globalization affect growth? {Evidence} from a new index of globalization}.{\BBCQ}
\newblock
\APACjournalVolNumPages{Applied Economics}{38}{10}{1091--1110}.
\PrintBackRefs{\CurrentBib}

\bibitem [\protect \citeauthoryear {%
Fares%
, Chung%
\BCBL {}\ \BBA {} Abbasi%
}{%
Fares%
\ \protect \BOthers {.}}{%
{\protect \APACyear {2021}}%
}]{%
fares2021stakeholder}
\APACinsertmetastar {%
fares2021stakeholder}%
\begin{APACrefauthors}%
Fares, J.%
, Chung, K\BPBI S\BPBI K.%
\BCBL {}\ \BBA {} Abbasi, A.%
\end{APACrefauthors}%
\unskip\
\newblock
\APACrefYearMonthDay{2021}{}{}.
\newblock
{\BBOQ}\APACrefatitle {Stakeholder theory and management: Understanding longitudinal collaboration networks} {Stakeholder theory and management: Understanding longitudinal collaboration networks}.{\BBCQ}
\newblock
\APACjournalVolNumPages{PLOS ONE}{16}{10}{e0255658}.
\newblock
\begin{APACrefDOI} \doi{10.1371/journal.pone.0255658} \end{APACrefDOI}
\PrintBackRefs{\CurrentBib}

\bibitem [\protect \citeauthoryear {%
Ferligoj%
, Kronegger%
, Mali%
, Snijders%
\BCBL {}\ \BBA {} Doreian%
}{%
Ferligoj%
\ \protect \BOthers {.}}{%
{\protect \APACyear {2015}}%
}]{%
ferligoj2015scientific}
\APACinsertmetastar {%
ferligoj2015scientific}%
\begin{APACrefauthors}%
Ferligoj, A.%
, Kronegger, L.%
, Mali, F.%
, Snijders, T\BPBI A\BPBI B.%
\BCBL {}\ \BBA {} Doreian, P.%
\end{APACrefauthors}%
\unskip\
\newblock
\APACrefYearMonthDay{2015}{}{}.
\newblock
{\BBOQ}\APACrefatitle {Scientific collaboration dynamics in a national scientific system} {Scientific collaboration dynamics in a national scientific system}.{\BBCQ}
\newblock
\APACjournalVolNumPages{Scientometrics}{104}{3}{985--1012}.
\PrintBackRefs{\CurrentBib}

\bibitem [\protect \citeauthoryear {%
Fernandez%
, Ferr{\'a}ndiz%
\BCBL {}\ \BBA {} Le{\'o}n%
}{%
Fernandez%
\ \protect \BOthers {.}}{%
{\protect \APACyear {2016}}%
}]{%
fernandez2016proximity}
\APACinsertmetastar {%
fernandez2016proximity}%
\begin{APACrefauthors}%
Fernandez, A.%
, Ferr{\'a}ndiz, E.%
\BCBL {}\ \BBA {} Le{\'o}n, M\BPBI D.%
\end{APACrefauthors}%
\unskip\
\newblock
\APACrefYearMonthDay{2016}{}{}.
\newblock
{\BBOQ}\APACrefatitle {Proximity dimensions and scientific collaboration among academic institutions in {Europe}: The closer, the better?} {Proximity dimensions and scientific collaboration among academic institutions in {Europe}: The closer, the better?}{\BBCQ}
\newblock
\APACjournalVolNumPages{Scientometrics}{106}{3}{1073--1092}.
\PrintBackRefs{\CurrentBib}

\bibitem [\protect \citeauthoryear {%
Fortunato%
\ \protect \BOthers {.}}{%
Fortunato%
\ \protect \BOthers {.}}{%
{\protect \APACyear {2018}}%
}]{%
fortunato2018science}
\APACinsertmetastar {%
fortunato2018science}%
\begin{APACrefauthors}%
Fortunato, S.%
, Bergstrom, C\BPBI T.%
, B{\"o}rner, K.%
, Evans, J\BPBI A.%
, Helbing, D.%
, Milojevi{\'c}, S.%
\BDBL {}others%
\end{APACrefauthors}%
\unskip\
\newblock
\APACrefYearMonthDay{2018}{}{}.
\newblock
{\BBOQ}\APACrefatitle {Science of science} {Science of science}.{\BBCQ}
\newblock
\APACjournalVolNumPages{Science}{359}{6379}{eaao0185}.
\PrintBackRefs{\CurrentBib}

\bibitem [\protect \citeauthoryear {%
Frame%
\ \BBA {} Carpenter%
}{%
Frame%
\ \BBA {} Carpenter%
}{%
{\protect \APACyear {1979}}%
}]{%
frame1979international}
\APACinsertmetastar {%
frame1979international}%
\begin{APACrefauthors}%
Frame, J\BPBI D.%
\BCBT {}\ \BBA {} Carpenter, M\BPBI P.%
\end{APACrefauthors}%
\unskip\
\newblock
\APACrefYearMonthDay{1979}{}{}.
\newblock
{\BBOQ}\APACrefatitle {International Research Collaboration} {International research collaboration}.{\BBCQ}
\newblock
\APACjournalVolNumPages{Social Studies of Science}{9}{4}{481--497}.
\newblock
\begin{APACrefDOI} \doi{10.1177/030631277900900404} \end{APACrefDOI}
\PrintBackRefs{\CurrentBib}

\bibitem [\protect \citeauthoryear {%
Freeman%
}{%
Freeman%
}{%
{\protect \APACyear {1992}}%
}]{%
freeman1992}
\APACinsertmetastar {%
freeman1992}%
\begin{APACrefauthors}%
Freeman, C.%
\end{APACrefauthors}%
\unskip\
\newblock
\APACrefYearMonthDay{1992}{}{}.
\newblock
{\BBOQ}\APACrefatitle {Formal Scientific and Technical Institutions in the National System of Innovation} {Formal scientific and technical institutions in the national system of innovation}.{\BBCQ}
\newblock
\BIn{} B\BHBI {\AA}.~Lundvall\ (\BED), \APACrefbtitle {National Systems of Innovation: Toward a Theory of Innovation and Interactive Learning} {National systems of innovation: Toward a theory of innovation and interactive learning}\ (\BPGS\ 169--187).
\newblock
\APACaddressPublisher{London}{Pinter}.
\PrintBackRefs{\CurrentBib}

\bibitem [\protect \citeauthoryear {%
Frenken%
, Hardeman%
\BCBL {}\ \BBA {} Hoekman%
}{%
Frenken%
\ \protect \BOthers {.}%
}{%
{\protect \APACyear {2009}}%
}]{%
frenken2009spatial}
\APACinsertmetastar {%
frenken2009spatial}%
\begin{APACrefauthors}%
Frenken, K.%
, Hardeman, S.%
\BCBL {}\ \BBA {} Hoekman, J.%
\end{APACrefauthors}%
\unskip\
\newblock
\APACrefYearMonthDay{2009}{}{}.
\newblock
{\BBOQ}\APACrefatitle {Spatial scientometrics: Towards a cumulative research program} {Spatial scientometrics: Towards a cumulative research program}.{\BBCQ}
\newblock
\APACjournalVolNumPages{Journal of Informetrics}{3}{3}{222--232}.
\PrintBackRefs{\CurrentBib}

\bibitem [\protect \citeauthoryear {%
Fu%
, Marques%
, Tseng%
, Powell%
\BCBL {}\ \BBA {} Baker%
}{%
Fu%
\ \protect \BOthers {.}}{%
{\protect \APACyear {2022}}%
}]{%
fu2022evolving}
\APACinsertmetastar {%
fu2022evolving}%
\begin{APACrefauthors}%
Fu, Y\BPBI C.%
, Marques, M.%
, Tseng, Y\BHBI H.%
, Powell, J\BPBI J\BPBI W.%
\BCBL {}\ \BBA {} Baker, D\BPBI P.%
\end{APACrefauthors}%
\unskip\
\newblock
\APACrefYearMonthDay{2022}{}{}.
\newblock
{\BBOQ}\APACrefatitle {An evolving international research collaboration network: Spatial and thematic developments in co-authored higher education research, 1998--2018} {An evolving international research collaboration network: Spatial and thematic developments in co-authored higher education research, 1998--2018}.{\BBCQ}
\newblock
\APACjournalVolNumPages{Scientometrics}{127}{3}{1403--1429}.
\PrintBackRefs{\CurrentBib}

\bibitem [\protect \citeauthoryear {%
Garfield%
}{%
Garfield%
}{%
{\protect \APACyear {1979}}%
}]{%
garfield1979citation}
\APACinsertmetastar {%
garfield1979citation}%
\begin{APACrefauthors}%
Garfield, E.%
\end{APACrefauthors}%
\unskip\
\newblock
\APACrefYearMonthDay{1979}{}{}.
\newblock
{\BBOQ}\APACrefatitle {Is citation analysis a legitimate evaluation tool?} {Is citation analysis a legitimate evaluation tool?}{\BBCQ}
\newblock
\APACjournalVolNumPages{Scientometrics}{1}{}{359--375}.
\PrintBackRefs{\CurrentBib}

\bibitem [\protect \citeauthoryear {%
Gl{\"a}nzel%
}{%
Gl{\"a}nzel%
}{%
{\protect \APACyear {2001}}%
}]{%
glanzel2001national}
\APACinsertmetastar {%
glanzel2001national}%
\begin{APACrefauthors}%
Gl{\"a}nzel, W.%
\end{APACrefauthors}%
\unskip\
\newblock
\APACrefYearMonthDay{2001}{}{}.
\newblock
{\BBOQ}\APACrefatitle {National characteristics in international scientific co-authorship relations} {National characteristics in international scientific co-authorship relations}.{\BBCQ}
\newblock
\APACjournalVolNumPages{Scientometrics}{51}{}{69--115}.
\PrintBackRefs{\CurrentBib}

\bibitem [\protect \citeauthoryear {%
Gl{\"a}nzel%
\ \BBA {} De~Lange%
}{%
Gl{\"a}nzel%
\ \BBA {} De~Lange%
}{%
{\protect \APACyear {2002}}%
}]{%
glanzel2002distributional}
\APACinsertmetastar {%
glanzel2002distributional}%
\begin{APACrefauthors}%
Gl{\"a}nzel, W.%
\BCBT {}\ \BBA {} De~Lange, C.%
\end{APACrefauthors}%
\unskip\
\newblock
\APACrefYearMonthDay{2002}{}{}.
\newblock
{\BBOQ}\APACrefatitle {A distributional approach to multinationality measures of international scientific collaboration} {A distributional approach to multinationality measures of international scientific collaboration}.{\BBCQ}
\newblock
\APACjournalVolNumPages{Scientometrics}{54}{1}{75--89}.
\PrintBackRefs{\CurrentBib}

\bibitem [\protect \citeauthoryear {%
Gl{\"a}nzel%
\ \BBA {} Schubert%
}{%
Gl{\"a}nzel%
\ \BBA {} Schubert%
}{%
{\protect \APACyear {2004}}%
}]{%
glanzel2004analysing}
\APACinsertmetastar {%
glanzel2004analysing}%
\begin{APACrefauthors}%
Gl{\"a}nzel, W.%
\BCBT {}\ \BBA {} Schubert, A.%
\end{APACrefauthors}%
\unskip\
\newblock
\APACrefYearMonthDay{2004}{}{}.
\newblock
{\BBOQ}\APACrefatitle {Analysing scientific networks through co-authorship} {Analysing scientific networks through co-authorship}.{\BBCQ}
\newblock
\BIn{} \APACrefbtitle {Handbook of Quantitative Science and Technology Research: The Use of Publication and Patent Statistics in Studies of {S\&T} Systems} {Handbook of quantitative science and technology research: The use of publication and patent statistics in studies of {S\&T} systems}\ (\BPGS\ 257--276).
\newblock
\APACaddressPublisher{}{Springer}.
\PrintBackRefs{\CurrentBib}

\bibitem [\protect \citeauthoryear {%
Gomez%
, Herman%
\BCBL {}\ \BBA {} Parigi%
}{%
Gomez%
\ \protect \BOthers {.}}{%
{\protect \APACyear {2022}}%
}]{%
gomez2022leading}
\APACinsertmetastar {%
gomez2022leading}%
\begin{APACrefauthors}%
Gomez, C\BPBI J.%
, Herman, A\BPBI C.%
\BCBL {}\ \BBA {} Parigi, P.%
\end{APACrefauthors}%
\unskip\
\newblock
\APACrefYearMonthDay{2022}{}{}.
\newblock
{\BBOQ}\APACrefatitle {Leading countries in global science increasingly receive more citations than other countries doing similar research} {Leading countries in global science increasingly receive more citations than other countries doing similar research}.{\BBCQ}
\newblock
\APACjournalVolNumPages{Nature Human Behaviour}{6}{7}{919--929}.
\PrintBackRefs{\CurrentBib}

\bibitem [\protect \citeauthoryear {%
Granovetter%
}{%
Granovetter%
}{%
{\protect \APACyear {1985}}%
}]{%
granovetter1985economic}
\APACinsertmetastar {%
granovetter1985economic}%
\begin{APACrefauthors}%
Granovetter, M.%
\end{APACrefauthors}%
\unskip\
\newblock
\APACrefYearMonthDay{1985}{}{}.
\newblock
{\BBOQ}\APACrefatitle {Economic action and social structure: The problem of embeddedness} {Economic action and social structure: The problem of embeddedness}.{\BBCQ}
\newblock
\APACjournalVolNumPages{American Journal of Sociology}{91}{3}{481--510}.
\PrintBackRefs{\CurrentBib}

\bibitem [\protect \citeauthoryear {%
Gui%
, Liu%
\BCBL {}\ \BBA {} Du%
}{%
Gui%
\ \protect \BOthers {.}}{%
{\protect \APACyear {2019}}%
}]{%
gui2019globalization}
\APACinsertmetastar {%
gui2019globalization}%
\begin{APACrefauthors}%
Gui, Q.%
, Liu, C.%
\BCBL {}\ \BBA {} Du, D.%
\end{APACrefauthors}%
\unskip\
\newblock
\APACrefYearMonthDay{2019}{}{}.
\newblock
{\BBOQ}\APACrefatitle {Globalization of science and international scientific collaboration: A network perspective} {Globalization of science and international scientific collaboration: A network perspective}.{\BBCQ}
\newblock
\APACjournalVolNumPages{Geoforum}{105}{}{1--12}.
\PrintBackRefs{\CurrentBib}

\bibitem [\protect \citeauthoryear {%
Gygli%
, Haelg%
, Potrafke%
\BCBL {}\ \BBA {} Sturm%
}{%
Gygli%
\ \protect \BOthers {.}%
}{%
{\protect \APACyear {2019}}%
}]{%
gygli2019kof}
\APACinsertmetastar {%
gygli2019kof}%
\begin{APACrefauthors}%
Gygli, S.%
, Haelg, F.%
, Potrafke, N.%
\BCBL {}\ \BBA {} Sturm, J.-E.%
\end{APACrefauthors}%
\unskip\
\newblock
\APACrefYearMonthDay{2019}{}{}.
\newblock
{\BBOQ}\APACrefatitle {The {KOF} Globalisation Index -- revisited} {The {KOF} Globalisation Index -- revisited}.{\BBCQ}
\newblock
\APACjournalVolNumPages{The Review of International Organizations}{14}{3}{543--574}.
\PrintBackRefs{\CurrentBib}

\bibitem [\protect \citeauthoryear {%
Hoekman%
, Frenken%
\BCBL {}\ \BBA {} Tijssen%
}{%
Hoekman%
\ \protect \BOthers {.}%
}{%
{\protect \APACyear {2010}}%
}]{%
hoekman2010research}
\APACinsertmetastar {%
hoekman2010research}%
\begin{APACrefauthors}%
Hoekman, J.%
, Frenken, K.%
\BCBL {}\ \BBA {} Tijssen, R\BPBI J\BPBI W.%
\end{APACrefauthors}%
\unskip\
\newblock
\APACrefYearMonthDay{2010}{}{}.
\newblock
{\BBOQ}\APACrefatitle {Research collaboration at a distance: Changing spatial patterns of scientific collaboration within {Europe}} {Research collaboration at a distance: Changing spatial patterns of scientific collaboration within {Europe}}.{\BBCQ}
\newblock
\APACjournalVolNumPages{Research Policy}{39}{5}{662--673}.
\PrintBackRefs{\CurrentBib}

\bibitem [\protect \citeauthoryear {%
Hou%
, Pan%
\BCBL {}\ \BBA {} Zhu%
}{%
Hou%
\ \protect \BOthers {.}}{%
{\protect \APACyear {2021}}%
}]{%
hou2021impact}
\APACinsertmetastar {%
hou2021impact}%
\begin{APACrefauthors}%
Hou, L.%
, Pan, Y.%
\BCBL {}\ \BBA {} Zhu, J\BPBI J\BPBI H.%
\end{APACrefauthors}%
\unskip\
\newblock
\APACrefYearMonthDay{2021}{}{}.
\newblock
{\BBOQ}\APACrefatitle {Impact of scientific, economic, geopolitical, and cultural factors on international research collaboration} {Impact of scientific, economic, geopolitical, and cultural factors on international research collaboration}.{\BBCQ}
\newblock
\APACjournalVolNumPages{Journal of Informetrics}{15}{3}{101194}.
\PrintBackRefs{\CurrentBib}

\bibitem [\protect \citeauthoryear {%
Huisman%
\ \BBA {} Steglich%
}{%
Huisman%
\ \BBA {} Steglich%
}{%
{\protect \APACyear {2008}}%
}]{%
huisman2008treatment}
\APACinsertmetastar {%
huisman2008treatment}%
\begin{APACrefauthors}%
Huisman, M.%
\BCBT {}\ \BBA {} Steglich, C.%
\end{APACrefauthors}%
\unskip\
\newblock
\APACrefYearMonthDay{2008}{}{}.
\newblock
{\BBOQ}\APACrefatitle {Treatment of non-response in longitudinal network studies} {Treatment of non-response in longitudinal network studies}.{\BBCQ}
\newblock
\APACjournalVolNumPages{Social Networks}{30}{4}{297--308}.
\PrintBackRefs{\CurrentBib}

\bibitem [\protect \citeauthoryear {%
Hwang%
}{%
Hwang%
}{%
{\protect \APACyear {2005}}%
}]{%
hwang2005inferior}
\APACinsertmetastar {%
hwang2005inferior}%
\begin{APACrefauthors}%
Hwang, K.%
\end{APACrefauthors}%
\unskip\
\newblock
\APACrefYearMonthDay{2005}{}{}.
\newblock
{\BBOQ}\APACrefatitle {The inferior science and the dominant use of {English} in knowledge production: A case study of {Korean} science and technology} {The inferior science and the dominant use of {English} in knowledge production: A case study of {Korean} science and technology}.{\BBCQ}
\newblock
\APACjournalVolNumPages{Science Communication}{26}{4}{390--427}.
\PrintBackRefs{\CurrentBib}

\bibitem [\protect \citeauthoryear {%
Hwang%
}{%
Hwang%
}{%
{\protect \APACyear {2013}}%
}]{%
hwang2013effects}
\APACinsertmetastar {%
hwang2013effects}%
\begin{APACrefauthors}%
Hwang, K.%
\end{APACrefauthors}%
\unskip\
\newblock
\APACrefYearMonthDay{2013}{}{}.
\newblock
{\BBOQ}\APACrefatitle {Effects of the language barrier on processes and performance of international scientific collaboration, collaborators' participation, organizational integrity, and interorganizational relationships} {Effects of the language barrier on processes and performance of international scientific collaboration, collaborators' participation, organizational integrity, and interorganizational relationships}.{\BBCQ}
\newblock
\APACjournalVolNumPages{Science Communication}{35}{1}{3--31}.
\PrintBackRefs{\CurrentBib}

\bibitem [\protect \citeauthoryear {%
Jeong%
, N{\'e}da%
\BCBL {}\ \BBA {} Barab{\'a}si%
}{%
Jeong%
\ \protect \BOthers {.}}{%
{\protect \APACyear {2003}}%
}]{%
jeong2003measuring}
\APACinsertmetastar {%
jeong2003measuring}%
\begin{APACrefauthors}%
Jeong, H.%
, N{\'e}da, Z.%
\BCBL {}\ \BBA {} Barab{\'a}si, A\BHBI L.%
\end{APACrefauthors}%
\unskip\
\newblock
\APACrefYearMonthDay{2003}{}{}.
\newblock
{\BBOQ}\APACrefatitle {Measuring preferential attachment in evolving networks} {Measuring preferential attachment in evolving networks}.{\BBCQ}
\newblock
\APACjournalVolNumPages{Europhysics Letters}{61}{4}{567}.
\PrintBackRefs{\CurrentBib}

\bibitem [\protect \citeauthoryear {%
Jha%
\ \BBA {} Welch%
}{%
Jha%
\ \BBA {} Welch%
}{%
{\protect \APACyear {2010}}%
}]{%
jha2010relational}
\APACinsertmetastar {%
jha2010relational}%
\begin{APACrefauthors}%
Jha, Y.%
\BCBT {}\ \BBA {} Welch, E\BPBI W.%
\end{APACrefauthors}%
\unskip\
\newblock
\APACrefYearMonthDay{2010}{}{}.
\newblock
{\BBOQ}\APACrefatitle {Relational mechanisms governing multifaceted collaborative behavior of academic scientists in six fields of science and engineering} {Relational mechanisms governing multifaceted collaborative behavior of academic scientists in six fields of science and engineering}.{\BBCQ}
\newblock
\APACjournalVolNumPages{Research Policy}{39}{9}{1174--1184}.
\PrintBackRefs{\CurrentBib}

\bibitem [\protect \citeauthoryear {%
Katz%
}{%
Katz%
}{%
{\protect \APACyear {1994}}%
}]{%
katz1994geographical}
\APACinsertmetastar {%
katz1994geographical}%
\begin{APACrefauthors}%
Katz, J\BPBI S.%
\end{APACrefauthors}%
\unskip\
\newblock
\APACrefYearMonthDay{1994}{}{}.
\newblock
{\BBOQ}\APACrefatitle {Geographical proximity and scientific collaboration} {Geographical proximity and scientific collaboration}.{\BBCQ}
\newblock
\APACjournalVolNumPages{Scientometrics}{31}{1}{31--43}.
\PrintBackRefs{\CurrentBib}

\bibitem [\protect \citeauthoryear {%
Katz%
}{%
Katz%
}{%
{\protect \APACyear {2016}}%
}]{%
katz2016complex}
\APACinsertmetastar {%
katz2016complex}%
\begin{APACrefauthors}%
Katz, J\BPBI S.%
\end{APACrefauthors}%
\unskip\
\newblock
\APACrefYearMonthDay{2016}{}{}.
\newblock
{\BBOQ}\APACrefatitle {What is a complex innovation system?} {What is a complex innovation system?}{\BBCQ}
\newblock
\APACjournalVolNumPages{PLOS ONE}{11}{6}{e0156150}.
\PrintBackRefs{\CurrentBib}

\bibitem [\protect \citeauthoryear {%
Katz%
\ \BBA {} Martin%
}{%
Katz%
\ \BBA {} Martin%
}{%
{\protect \APACyear {1997}}%
}]{%
katz1997research}
\APACinsertmetastar {%
katz1997research}%
\begin{APACrefauthors}%
Katz, J\BPBI S.%
\BCBT {}\ \BBA {} Martin, B\BPBI R.%
\end{APACrefauthors}%
\unskip\
\newblock
\APACrefYearMonthDay{1997}{}{}.
\newblock
{\BBOQ}\APACrefatitle {What is research collaboration?} {What is research collaboration?}{\BBCQ}
\newblock
\APACjournalVolNumPages{Research Policy}{26}{1}{1--18}.
\PrintBackRefs{\CurrentBib}

\bibitem [\protect \citeauthoryear {%
King%
}{%
King%
}{%
{\protect \APACyear {2004}}%
}]{%
king2004scientific}
\APACinsertmetastar {%
king2004scientific}%
\begin{APACrefauthors}%
King, D\BPBI A.%
\end{APACrefauthors}%
\unskip\
\newblock
\APACrefYearMonthDay{2004}{}{}.
\newblock
{\BBOQ}\APACrefatitle {The scientific impact of nations} {The scientific impact of nations}.{\BBCQ}
\newblock
\APACjournalVolNumPages{Nature}{430}{6997}{311--316}.
\PrintBackRefs{\CurrentBib}

\bibitem [\protect \citeauthoryear {%
Koka%
\ \BBA {} Prescott%
}{%
Koka%
\ \BBA {} Prescott%
}{%
{\protect \APACyear {2008}}%
}]{%
koka2008designing}
\APACinsertmetastar {%
koka2008designing}%
\begin{APACrefauthors}%
Koka, B\BPBI R.%
\BCBT {}\ \BBA {} Prescott, J\BPBI E.%
\end{APACrefauthors}%
\unskip\
\newblock
\APACrefYearMonthDay{2008}{}{}.
\newblock
{\BBOQ}\APACrefatitle {Designing alliance networks: the influence of network position, environmental change, and strategy on firm performance} {Designing alliance networks: the influence of network position, environmental change, and strategy on firm performance}.{\BBCQ}
\newblock
\APACjournalVolNumPages{Strategic Management Journal}{29}{6}{639--661}.
\PrintBackRefs{\CurrentBib}

\bibitem [\protect \citeauthoryear {%
Kuzhabekova%
, Hendel%
\BCBL {}\ \BBA {} Chapman%
}{%
Kuzhabekova%
\ \protect \BOthers {.}}{%
{\protect \APACyear {2015}}%
}]{%
kuzhabekova2015mapping}
\APACinsertmetastar {%
kuzhabekova2015mapping}%
\begin{APACrefauthors}%
Kuzhabekova, A.%
, Hendel, D\BPBI D.%
\BCBL {}\ \BBA {} Chapman, D\BPBI W.%
\end{APACrefauthors}%
\unskip\
\newblock
\APACrefYearMonthDay{2015}{}{}.
\newblock
{\BBOQ}\APACrefatitle {Mapping global research on international higher education} {Mapping global research on international higher education}.{\BBCQ}
\newblock
\APACjournalVolNumPages{Research in Higher Education}{56}{}{861--882}.
\PrintBackRefs{\CurrentBib}

\bibitem [\protect \citeauthoryear {%
Kwiek%
}{%
Kwiek%
}{%
{\protect \APACyear {2021}}%
}]{%
kwiek2021large}
\APACinsertmetastar {%
kwiek2021large}%
\begin{APACrefauthors}%
Kwiek, M.%
\end{APACrefauthors}%
\unskip\
\newblock
\APACrefYearMonthDay{2021}{}{}.
\newblock
{\BBOQ}\APACrefatitle {What large-scale publication and citation data tell us about international research collaboration in {Europe}: Changing national patterns in global contexts} {What large-scale publication and citation data tell us about international research collaboration in {Europe}: Changing national patterns in global contexts}.{\BBCQ}
\newblock
\APACjournalVolNumPages{Studies in Higher Education}{46}{12}{2629--2649}.
\PrintBackRefs{\CurrentBib}

\bibitem [\protect \citeauthoryear {%
Leahey%
}{%
Leahey%
}{%
{\protect \APACyear {2016}}%
}]{%
leahey_sole_2016}
\APACinsertmetastar {%
leahey_sole_2016}%
\begin{APACrefauthors}%
Leahey, E.%
\end{APACrefauthors}%
\unskip\
\newblock
\APACrefYearMonthDay{2016}{}{}.
\newblock
{\BBOQ}\APACrefatitle {From Sole Investigator to Team Scientist: Trends in the Practice and Study of Research Collaboration} {From sole investigator to team scientist: Trends in the practice and study of research collaboration}.{\BBCQ}
\newblock
\APACjournalVolNumPages{Annual Review of Sociology}{42}{}{81--100}.
\newblock
\begin{APACrefDOI} \doi{10.1146/annurev-soc-081715-074219} \end{APACrefDOI}
\PrintBackRefs{\CurrentBib}

\bibitem [\protect \citeauthoryear {%
Leydesdorff%
\ \BBA {} Wagner%
}{%
Leydesdorff%
\ \BBA {} Wagner%
}{%
{\protect \APACyear {2008}}%
}]{%
leydesdorff2008international}
\APACinsertmetastar {%
leydesdorff2008international}%
\begin{APACrefauthors}%
Leydesdorff, L.%
\BCBT {}\ \BBA {} Wagner, C\BPBI S.%
\end{APACrefauthors}%
\unskip\
\newblock
\APACrefYearMonthDay{2008}{}{}.
\newblock
{\BBOQ}\APACrefatitle {International collaboration in science and the formation of a core group} {International collaboration in science and the formation of a core group}.{\BBCQ}
\newblock
\APACjournalVolNumPages{Journal of Informetrics}{2}{4}{317--325}.
\PrintBackRefs{\CurrentBib}

\bibitem [\protect \citeauthoryear {%
E\BPBI Y.~Li%
, Liao%
\BCBL {}\ \BBA {} Yen%
}{%
E\BPBI Y.~Li%
\ \protect \BOthers {.}}{%
{\protect \APACyear {2013}}%
}]{%
li2013co}
\APACinsertmetastar {%
li2013co}%
\begin{APACrefauthors}%
Li, E\BPBI Y.%
, Liao, C\BPBI H.%
\BCBL {}\ \BBA {} Yen, H\BPBI R.%
\end{APACrefauthors}%
\unskip\
\newblock
\APACrefYearMonthDay{2013}{}{}.
\newblock
{\BBOQ}\APACrefatitle {Co-authorship networks and research impact: A social capital perspective} {Co-authorship networks and research impact: A social capital perspective}.{\BBCQ}
\newblock
\APACjournalVolNumPages{Research Policy}{42}{9}{1515--1530}.
\PrintBackRefs{\CurrentBib}

\bibitem [\protect \citeauthoryear {%
W.~Li%
, Zhang%
, Zheng%
, Cranmer%
\BCBL {}\ \BBA {} Clauset%
}{%
W.~Li%
\ \protect \BOthers {.}}{%
{\protect \APACyear {2022}}%
}]{%
li2022untangling}
\APACinsertmetastar {%
li2022untangling}%
\begin{APACrefauthors}%
Li, W.%
, Zhang, S.%
, Zheng, Z.%
, Cranmer, S\BPBI J.%
\BCBL {}\ \BBA {} Clauset, A.%
\end{APACrefauthors}%
\unskip\
\newblock
\APACrefYearMonthDay{2022}{}{}.
\newblock
{\BBOQ}\APACrefatitle {Untangling the network effects of productivity and prominence among scientists} {Untangling the network effects of productivity and prominence among scientists}.{\BBCQ}
\newblock
\APACjournalVolNumPages{Nature Communications}{13}{1}{4907}.
\newblock
\begin{APACrefDOI} \doi{10.1038/s41467-022-32604-6} \end{APACrefDOI}
\PrintBackRefs{\CurrentBib}

\bibitem [\protect \citeauthoryear {%
Liu%
, Zhang%
, Zhang%
\BCBL {}\ \BBA {} You%
}{%
Liu%
\ \protect \BOthers {.}}{%
{\protect \APACyear {2022}}%
}]{%
liu2022scientific}
\APACinsertmetastar {%
liu2022scientific}%
\begin{APACrefauthors}%
Liu, Y.%
, Zhang, M.%
, Zhang, G.%
\BCBL {}\ \BBA {} You, X.%
\end{APACrefauthors}%
\unskip\
\newblock
\APACrefYearMonthDay{2022}{}{}.
\newblock
{\BBOQ}\APACrefatitle {Scientific elites versus other scientists: who are better at taking advantage of the research collaboration network?} {Scientific elites versus other scientists: who are better at taking advantage of the research collaboration network?}{\BBCQ}
\newblock
\APACjournalVolNumPages{Scientometrics}{127}{6}{3145--3166}.
\PrintBackRefs{\CurrentBib}

\bibitem [\protect \citeauthoryear {%
Lospinoso%
\ \BBA {} Snijders%
}{%
Lospinoso%
\ \BBA {} Snijders%
}{%
{\protect \APACyear {2019}}%
}]{%
lospinoso2019goodness}
\APACinsertmetastar {%
lospinoso2019goodness}%
\begin{APACrefauthors}%
Lospinoso, J.%
\BCBT {}\ \BBA {} Snijders, T\BPBI A\BPBI B.%
\end{APACrefauthors}%
\unskip\
\newblock
\APACrefYearMonthDay{2019}{}{}.
\newblock
{\BBOQ}\APACrefatitle {Goodness of fit for stochastic actor-oriented models} {Goodness of fit for stochastic actor-oriented models}.{\BBCQ}
\newblock
\APACjournalVolNumPages{Methodological Innovations}{12}{3}{2059799119884282}.
\PrintBackRefs{\CurrentBib}

\bibitem [\protect \citeauthoryear {%
Luukkonen%
, Persson%
\BCBL {}\ \BBA {} Sivertsen%
}{%
Luukkonen%
\ \protect \BOthers {.}}{%
{\protect \APACyear {1992}}%
}]{%
luukkonen1992understanding}
\APACinsertmetastar {%
luukkonen1992understanding}%
\begin{APACrefauthors}%
Luukkonen, T.%
, Persson, O.%
\BCBL {}\ \BBA {} Sivertsen, G.%
\end{APACrefauthors}%
\unskip\
\newblock
\APACrefYearMonthDay{1992}{}{}.
\newblock
{\BBOQ}\APACrefatitle {Understanding patterns of international scientific collaboration} {Understanding patterns of international scientific collaboration}.{\BBCQ}
\newblock
\APACjournalVolNumPages{Science, Technology, \& Human Values}{17}{1}{101--126}.
\PrintBackRefs{\CurrentBib}

\bibitem [\protect \citeauthoryear {%
Luukkonen%
, Tijssen%
, Persson%
\BCBL {}\ \BBA {} Sivertsen%
}{%
Luukkonen%
\ \protect \BOthers {.}}{%
{\protect \APACyear {1993}}%
}]{%
luukkonen1993measurement}
\APACinsertmetastar {%
luukkonen1993measurement}%
\begin{APACrefauthors}%
Luukkonen, T.%
, Tijssen, R\BPBI J\BPBI W.%
, Persson, O.%
\BCBL {}\ \BBA {} Sivertsen, G.%
\end{APACrefauthors}%
\unskip\
\newblock
\APACrefYearMonthDay{1993}{}{}.
\newblock
{\BBOQ}\APACrefatitle {The measurement of international scientific collaboration} {The measurement of international scientific collaboration}.{\BBCQ}
\newblock
\APACjournalVolNumPages{Scientometrics}{28}{}{15--36}.
\PrintBackRefs{\CurrentBib}

\bibitem [\protect \citeauthoryear {%
Mallik%
\ \BBA {} Mandal%
}{%
Mallik%
\ \BBA {} Mandal%
}{%
{\protect \APACyear {2014}}%
}]{%
mallik2014bibliometric}
\APACinsertmetastar {%
mallik2014bibliometric}%
\begin{APACrefauthors}%
Mallik, A.%
\BCBT {}\ \BBA {} Mandal, N.%
\end{APACrefauthors}%
\unskip\
\newblock
\APACrefYearMonthDay{2014}{}{}.
\newblock
{\BBOQ}\APACrefatitle {Bibliometric analysis of global publication output and collaboration structure study in {microRNA} research} {Bibliometric analysis of global publication output and collaboration structure study in {microRNA} research}.{\BBCQ}
\newblock
\APACjournalVolNumPages{Scientometrics}{98}{}{2011--2037}.
\PrintBackRefs{\CurrentBib}

\bibitem [\protect \citeauthoryear {%
Melin%
\ \BBA {} Persson%
}{%
Melin%
\ \BBA {} Persson%
}{%
{\protect \APACyear {1996}}%
}]{%
melin1996studying}
\APACinsertmetastar {%
melin1996studying}%
\begin{APACrefauthors}%
Melin, G.%
\BCBT {}\ \BBA {} Persson, O.%
\end{APACrefauthors}%
\unskip\
\newblock
\APACrefYearMonthDay{1996}{}{}.
\newblock
{\BBOQ}\APACrefatitle {Studying research collaboration using co-authorships} {Studying research collaboration using co-authorships}.{\BBCQ}
\newblock
\APACjournalVolNumPages{Scientometrics}{36}{}{363--377}.
\PrintBackRefs{\CurrentBib}

\bibitem [\protect \citeauthoryear {%
Melitz%
\ \BBA {} Toubal%
}{%
Melitz%
\ \BBA {} Toubal%
}{%
{\protect \APACyear {2014}}%
}]{%
melitz2014native}
\APACinsertmetastar {%
melitz2014native}%
\begin{APACrefauthors}%
Melitz, J.%
\BCBT {}\ \BBA {} Toubal, F.%
\end{APACrefauthors}%
\unskip\
\newblock
\APACrefYearMonthDay{2014}{}{}.
\newblock
{\BBOQ}\APACrefatitle {Native language, spoken language, translation and trade} {Native language, spoken language, translation and trade}.{\BBCQ}
\newblock
\APACjournalVolNumPages{Journal of International Economics}{93}{2}{351--363}.
\PrintBackRefs{\CurrentBib}

\bibitem [\protect \citeauthoryear {%
Merton%
}{%
Merton%
}{%
{\protect \APACyear {1968}}%
}]{%
merton1968matthew}
\APACinsertmetastar {%
merton1968matthew}%
\begin{APACrefauthors}%
Merton, R\BPBI K.%
\end{APACrefauthors}%
\unskip\
\newblock
\APACrefYearMonthDay{1968}{}{}.
\newblock
{\BBOQ}\APACrefatitle {The {Matthew} effect in science: The reward and communication systems of science are considered} {The {Matthew} effect in science: The reward and communication systems of science are considered}.{\BBCQ}
\newblock
\APACjournalVolNumPages{Science}{159}{3810}{56--63}.
\PrintBackRefs{\CurrentBib}

\bibitem [\protect \citeauthoryear {%
Narin%
, Stevens%
\BCBL {}\ \BBA {} Whitlow%
}{%
Narin%
\ \protect \BOthers {.}}{%
{\protect \APACyear {1991}}%
}]{%
narin1991scientific}
\APACinsertmetastar {%
narin1991scientific}%
\begin{APACrefauthors}%
Narin, F.%
, Stevens, K.%
\BCBL {}\ \BBA {} Whitlow, E\BPBI S.%
\end{APACrefauthors}%
\unskip\
\newblock
\APACrefYearMonthDay{1991}{}{}.
\newblock
{\BBOQ}\APACrefatitle {Scientific co-operation in {Europe} and the citation of multinationally authored papers} {Scientific co-operation in {Europe} and the citation of multinationally authored papers}.{\BBCQ}
\newblock
\APACjournalVolNumPages{Scientometrics}{21}{}{313--323}.
\PrintBackRefs{\CurrentBib}

\bibitem [\protect \citeauthoryear {%
Neal%
}{%
Neal%
}{%
{\protect \APACyear {2014}}%
}]{%
neal2014backbone}
\APACinsertmetastar {%
neal2014backbone}%
\begin{APACrefauthors}%
Neal, Z\BPBI P.%
\end{APACrefauthors}%
\unskip\
\newblock
\APACrefYearMonthDay{2014}{}{}.
\newblock
{\BBOQ}\APACrefatitle {The backbone of bipartite projections: Inferring relationships from co-authorship, co-sponsorship, co-attendance and other co-behaviors} {The backbone of bipartite projections: Inferring relationships from co-authorship, co-sponsorship, co-attendance and other co-behaviors}.{\BBCQ}
\newblock
\APACjournalVolNumPages{Social Networks}{39}{}{84--97}.
\PrintBackRefs{\CurrentBib}

\bibitem [\protect \citeauthoryear {%
Neal%
}{%
Neal%
}{%
{\protect \APACyear {2022}}%
}]{%
neal2022backbone}
\APACinsertmetastar {%
neal2022backbone}%
\begin{APACrefauthors}%
Neal, Z\BPBI P.%
\end{APACrefauthors}%
\unskip\
\newblock
\APACrefYearMonthDay{2022}{}{}.
\newblock
{\BBOQ}\APACrefatitle {backbone: An {R} package to extract network backbones} {backbone: An {R} package to extract network backbones}.{\BBCQ}
\newblock
\APACjournalVolNumPages{PLOS ONE}{17}{5}{e0269137}.
\PrintBackRefs{\CurrentBib}

\bibitem [\protect \citeauthoryear {%
Newman%
}{%
Newman%
}{%
{\protect \APACyear {2001}}%
}]{%
newman2001structure}
\APACinsertmetastar {%
newman2001structure}%
\begin{APACrefauthors}%
Newman, M\BPBI E\BPBI J.%
\end{APACrefauthors}%
\unskip\
\newblock
\APACrefYearMonthDay{2001}{}{}.
\newblock
{\BBOQ}\APACrefatitle {The structure of scientific collaboration networks} {The structure of scientific collaboration networks}.{\BBCQ}
\newblock
\APACjournalVolNumPages{Proceedings of the National Academy of Sciences}{98}{2}{404--409}.
\PrintBackRefs{\CurrentBib}

\bibitem [\protect \citeauthoryear {%
Newman%
}{%
Newman%
}{%
{\protect \APACyear {2002}}%
}]{%
newman2002assortative}
\APACinsertmetastar {%
newman2002assortative}%
\begin{APACrefauthors}%
Newman, M\BPBI E\BPBI J.%
\end{APACrefauthors}%
\unskip\
\newblock
\APACrefYearMonthDay{2002}{}{}.
\newblock
{\BBOQ}\APACrefatitle {Assortative mixing in networks} {Assortative mixing in networks}.{\BBCQ}
\newblock
\APACjournalVolNumPages{Physical Review Letters}{89}{20}{208701}.
\PrintBackRefs{\CurrentBib}

\bibitem [\protect \citeauthoryear {%
Newman%
}{%
Newman%
}{%
{\protect \APACyear {2003}}%
}]{%
newman2003structure}
\APACinsertmetastar {%
newman2003structure}%
\begin{APACrefauthors}%
Newman, M\BPBI E\BPBI J.%
\end{APACrefauthors}%
\unskip\
\newblock
\APACrefYearMonthDay{2003}{}{}.
\newblock
{\BBOQ}\APACrefatitle {The structure and function of complex networks} {The structure and function of complex networks}.{\BBCQ}
\newblock
\APACjournalVolNumPages{SIAM Review}{45}{2}{167--256}.
\PrintBackRefs{\CurrentBib}

\bibitem [\protect \citeauthoryear {%
Newman%
}{%
Newman%
}{%
{\protect \APACyear {2004}}%
}]{%
newman2004coauthorship}
\APACinsertmetastar {%
newman2004coauthorship}%
\begin{APACrefauthors}%
Newman, M\BPBI E\BPBI J.%
\end{APACrefauthors}%
\unskip\
\newblock
\APACrefYearMonthDay{2004}{}{}.
\newblock
{\BBOQ}\APACrefatitle {Coauthorship networks and patterns of scientific collaboration} {Coauthorship networks and patterns of scientific collaboration}.{\BBCQ}
\newblock
\APACjournalVolNumPages{Proceedings of the National Academy of Sciences}{101}{suppl\_1}{5200--5205}.
\PrintBackRefs{\CurrentBib}

\bibitem [\protect \citeauthoryear {%
Nielsen%
\ \BBA {} Andersen%
}{%
Nielsen%
\ \BBA {} Andersen%
}{%
{\protect \APACyear {2021}}%
}]{%
nielsen2021global}
\APACinsertmetastar {%
nielsen2021global}%
\begin{APACrefauthors}%
Nielsen, M\BPBI W.%
\BCBT {}\ \BBA {} Andersen, J\BPBI P.%
\end{APACrefauthors}%
\unskip\
\newblock
\APACrefYearMonthDay{2021}{}{}.
\newblock
{\BBOQ}\APACrefatitle {Global citation inequality is on the rise} {Global citation inequality is on the rise}.{\BBCQ}
\newblock
\APACjournalVolNumPages{Proceedings of the National Academy of Sciences}{118}{7}{e2012208118}.
\PrintBackRefs{\CurrentBib}

\bibitem [\protect \citeauthoryear {%
Niezink%
, Snijders%
\BCBL {}\ \BBA {} van~Duijn%
}{%
Niezink%
\ \protect \BOthers {.}%
}{%
{\protect \APACyear {2019}}%
}]{%
niezink2019discrete}
\APACinsertmetastar {%
niezink2019discrete}%
\begin{APACrefauthors}%
Niezink, N\BPBI M\BPBI D.%
, Snijders, T\BPBI A\BPBI B.%
\BCBL {}\ \BBA {} van~Duijn, M\BPBI A\BPBI J.%
\end{APACrefauthors}%
\unskip\
\newblock
\APACrefYearMonthDay{2019}{}{}.
\newblock
{\BBOQ}\APACrefatitle {No longer discrete: Modeling the dynamics of social networks and continuous behavior} {No longer discrete: Modeling the dynamics of social networks and continuous behavior}.{\BBCQ}
\newblock
\APACjournalVolNumPages{Sociological Methodology}{49}{1}{295--340}.
\PrintBackRefs{\CurrentBib}

\bibitem [\protect \citeauthoryear {%
Nita%
}{%
Nita%
}{%
{\protect \APACyear {2019}}%
}]{%
nita2019empowering}
\APACinsertmetastar {%
nita2019empowering}%
\begin{APACrefauthors}%
Nita, A.%
\end{APACrefauthors}%
\unskip\
\newblock
\APACrefYearMonthDay{2019}{}{}.
\newblock
{\BBOQ}\APACrefatitle {Empowering impact assessments knowledge and international research collaboration---A bibliometric analysis of {Environmental Impact Assessment Review} journal} {Empowering impact assessments knowledge and international research collaboration---a bibliometric analysis of {Environmental Impact Assessment Review} journal}.{\BBCQ}
\newblock
\APACjournalVolNumPages{Environmental Impact Assessment Review}{78}{}{106283}.
\PrintBackRefs{\CurrentBib}

\bibitem [\protect \citeauthoryear {%
Niu%
\ \BBA {} Qiu%
}{%
Niu%
\ \BBA {} Qiu%
}{%
{\protect \APACyear {2014}}%
}]{%
niu2014network}
\APACinsertmetastar {%
niu2014network}%
\begin{APACrefauthors}%
Niu, F.%
\BCBT {}\ \BBA {} Qiu, J.%
\end{APACrefauthors}%
\unskip\
\newblock
\APACrefYearMonthDay{2014}{}{}.
\newblock
{\BBOQ}\APACrefatitle {Network structure, distribution and the growth of {Chinese} international research collaboration} {Network structure, distribution and the growth of {Chinese} international research collaboration}.{\BBCQ}
\newblock
\APACjournalVolNumPages{Scientometrics}{98}{}{1221--1233}.
\PrintBackRefs{\CurrentBib}

\bibitem [\protect \citeauthoryear {%
Olson%
\ \BBA {} Olson%
}{%
Olson%
\ \BBA {} Olson%
}{%
{\protect \APACyear {2000}}%
}]{%
olson2000distance}
\APACinsertmetastar {%
olson2000distance}%
\begin{APACrefauthors}%
Olson, G\BPBI M.%
\BCBT {}\ \BBA {} Olson, J\BPBI S.%
\end{APACrefauthors}%
\unskip\
\newblock
\APACrefYearMonthDay{2000}{}{}.
\newblock
{\BBOQ}\APACrefatitle {Distance matters} {Distance matters}.{\BBCQ}
\newblock
\APACjournalVolNumPages{Human-Computer Interaction}{15}{2-3}{139--178}.
\PrintBackRefs{\CurrentBib}

\bibitem [\protect \citeauthoryear {%
Onuchic%
\ \BBA {} Ray%
}{%
Onuchic%
\ \BBA {} Ray%
}{%
{\protect \APACyear {2023}}%
}]{%
onuchic2023signaling}
\APACinsertmetastar {%
onuchic2023signaling}%
\begin{APACrefauthors}%
Onuchic, P.%
\BCBT {}\ \BBA {} Ray, D.%
\end{APACrefauthors}%
\unskip\
\newblock
\APACrefYearMonthDay{2023}{}{}.
\newblock
{\BBOQ}\APACrefatitle {Signaling and discrimination in collaborative projects} {Signaling and discrimination in collaborative projects}.{\BBCQ}
\newblock
\APACjournalVolNumPages{American Economic Review}{113}{1}{210--252}.
\PrintBackRefs{\CurrentBib}

\bibitem [\protect \citeauthoryear {%
Ozmel%
, Reuer%
\BCBL {}\ \BBA {} Gulati%
}{%
Ozmel%
\ \protect \BOthers {.}}{%
{\protect \APACyear {2013}}%
}]{%
ozmel2013signals}
\APACinsertmetastar {%
ozmel2013signals}%
\begin{APACrefauthors}%
Ozmel, U.%
, Reuer, J\BPBI J.%
\BCBL {}\ \BBA {} Gulati, R.%
\end{APACrefauthors}%
\unskip\
\newblock
\APACrefYearMonthDay{2013}{}{}.
\newblock
{\BBOQ}\APACrefatitle {Signals across multiple networks: How venture capital and alliance networks affect interorganizational collaboration} {Signals across multiple networks: How venture capital and alliance networks affect interorganizational collaboration}.{\BBCQ}
\newblock
\APACjournalVolNumPages{Academy of Management Journal}{56}{3}{852--866}.
\PrintBackRefs{\CurrentBib}

\bibitem [\protect \citeauthoryear {%
Perc%
}{%
Perc%
}{%
{\protect \APACyear {2014}}%
}]{%
perc2014matthew}
\APACinsertmetastar {%
perc2014matthew}%
\begin{APACrefauthors}%
Perc, M.%
\end{APACrefauthors}%
\unskip\
\newblock
\APACrefYearMonthDay{2014}{}{}.
\newblock
{\BBOQ}\APACrefatitle {The {Matthew} effect in empirical data} {The {Matthew} effect in empirical data}.{\BBCQ}
\newblock
\APACjournalVolNumPages{Journal of The Royal Society Interface}{11}{98}{20140378}.
\PrintBackRefs{\CurrentBib}

\bibitem [\protect \citeauthoryear {%
Persson%
}{%
Persson%
}{%
{\protect \APACyear {2010}}%
}]{%
persson2010highly}
\APACinsertmetastar {%
persson2010highly}%
\begin{APACrefauthors}%
Persson, O.%
\end{APACrefauthors}%
\unskip\
\newblock
\APACrefYearMonthDay{2010}{}{}.
\newblock
{\BBOQ}\APACrefatitle {Are highly cited papers more international?} {Are highly cited papers more international?}{\BBCQ}
\newblock
\APACjournalVolNumPages{Scientometrics}{83}{2}{397--401}.
\PrintBackRefs{\CurrentBib}

\bibitem [\protect \citeauthoryear {%
P{\l}oszaj%
, Celi{\'n}ska-Janowicz%
\BCBL {}\ \BBA {} Olechnicka%
}{%
P{\l}oszaj%
\ \protect \BOthers {.}}{%
{\protect \APACyear {2018}}%
}]{%
ploszaj2018core}
\APACinsertmetastar {%
ploszaj2018core}%
\begin{APACrefauthors}%
P{\l}oszaj, A.%
, Celi{\'n}ska-Janowicz, D.%
\BCBL {}\ \BBA {} Olechnicka, A.%
\end{APACrefauthors}%
\unskip\
\newblock
\APACrefYearMonthDay{2018}{}{}.
\newblock
{\BBOQ}\APACrefatitle {Core-periphery relations in international research collaboration} {Core-periphery relations in international research collaboration}.{\BBCQ}
\newblock
\BIn{} \APACrefbtitle {Geographies of the University} {Geographies of the university}\ (\BPGS\ 75--101).
\newblock
\APACaddressPublisher{}{Springer}.
\newblock
\begin{APACrefDOI} \doi{10.1007/978-3-319-75593-9_3} \end{APACrefDOI}
\PrintBackRefs{\CurrentBib}

\bibitem [\protect \citeauthoryear {%
Podolny%
}{%
Podolny%
}{%
{\protect \APACyear {2001}}%
}]{%
podolny2001networks}
\APACinsertmetastar {%
podolny2001networks}%
\begin{APACrefauthors}%
Podolny, J\BPBI M.%
\end{APACrefauthors}%
\unskip\
\newblock
\APACrefYearMonthDay{2001}{}{}.
\newblock
{\BBOQ}\APACrefatitle {Networks as the pipes and prisms of the market} {Networks as the pipes and prisms of the market}.{\BBCQ}
\newblock
\APACjournalVolNumPages{American Journal of Sociology}{107}{1}{33--60}.
\PrintBackRefs{\CurrentBib}

\bibitem [\protect \citeauthoryear {%
Powell%
, Koput%
, Smith-Doerr%
\BCBL {}\ \BBA {} Owen-Smith%
}{%
Powell%
\ \protect \BOthers {.}}{%
{\protect \APACyear {1999}}%
}]{%
powell1999network}
\APACinsertmetastar {%
powell1999network}%
\begin{APACrefauthors}%
Powell, W\BPBI W.%
, Koput, K\BPBI W.%
, Smith-Doerr, L.%
\BCBL {}\ \BBA {} Owen-Smith, J.%
\end{APACrefauthors}%
\unskip\
\newblock
\APACrefYearMonthDay{1999}{}{}.
\newblock
{\BBOQ}\APACrefatitle {Network position and firm performance: Organizational returns to collaboration in the biotechnology industry} {Network position and firm performance: Organizational returns to collaboration in the biotechnology industry}.{\BBCQ}
\newblock
\APACjournalVolNumPages{Research in the Sociology of Organizations}{16}{1}{129--159}.
\PrintBackRefs{\CurrentBib}

\bibitem [\protect \citeauthoryear {%
Price%
}{%
Price%
}{%
{\protect \APACyear {1965}}%
}]{%
price1965networks}
\APACinsertmetastar {%
price1965networks}%
\begin{APACrefauthors}%
Price, D\BPBI J\BPBI D\BPBI S.%
\end{APACrefauthors}%
\unskip\
\newblock
\APACrefYearMonthDay{1965}{}{}.
\newblock
{\BBOQ}\APACrefatitle {Networks of scientific papers: The pattern of bibliographic references indicates the nature of the scientific research front} {Networks of scientific papers: The pattern of bibliographic references indicates the nature of the scientific research front}.{\BBCQ}
\newblock
\APACjournalVolNumPages{Science}{149}{3683}{510--515}.
\PrintBackRefs{\CurrentBib}

\bibitem [\protect \citeauthoryear {%
Purkayastha%
, Palmaro%
, Falk-Krzesinski%
\BCBL {}\ \BBA {} Baas%
}{%
Purkayastha%
\ \protect \BOthers {.}}{%
{\protect \APACyear {2019}}%
}]{%
purkayastha2019comparison}
\APACinsertmetastar {%
purkayastha2019comparison}%
\begin{APACrefauthors}%
Purkayastha, A.%
, Palmaro, E.%
, Falk-Krzesinski, H\BPBI J.%
\BCBL {}\ \BBA {} Baas, J.%
\end{APACrefauthors}%
\unskip\
\newblock
\APACrefYearMonthDay{2019}{}{}.
\newblock
{\BBOQ}\APACrefatitle {Comparison of two article-level, field-independent citation metrics: {Field-Weighted Citation Impact (FWCI)} and {Relative Citation Ratio (RCR)}} {Comparison of two article-level, field-independent citation metrics: {Field-Weighted Citation Impact (FWCI)} and {Relative Citation Ratio (RCR)}}.{\BBCQ}
\newblock
\APACjournalVolNumPages{Journal of Informetrics}{13}{2}{635--642}.
\PrintBackRefs{\CurrentBib}

\bibitem [\protect \citeauthoryear {%
Razzaq%
\ \protect \BOthers {.}}{%
Razzaq%
\ \protect \BOthers {.}}{%
{\protect \APACyear {2022}}%
}]{%
razzaq2022research}
\APACinsertmetastar {%
razzaq2022research}%
\begin{APACrefauthors}%
Razzaq, S.%
, Malik, A\BPBI K.%
, Raza, B.%
, Khattak, H\BPBI A.%
, Zegarra, G\BPBI W\BPBI M.%
\BCBL {}\ \BBA {} Zelada, Y\BPBI D.%
\end{APACrefauthors}%
\unskip\
\newblock
\APACrefYearMonthDay{2022}{}{}.
\newblock
{\BBOQ}\APACrefatitle {Research Collaboration Influence Analysis Using Dynamic Co-authorship and Citation Networks} {Research collaboration influence analysis using dynamic co-authorship and citation networks}.{\BBCQ}
\newblock
\APACjournalVolNumPages{International Journal of Interactive Multimedia and Artificial Intelligence}{7}{3}{103--116}.
\PrintBackRefs{\CurrentBib}

\bibitem [\protect \citeauthoryear {%
Redner%
}{%
Redner%
}{%
{\protect \APACyear {1998}}%
}]{%
redner1998popular}
\APACinsertmetastar {%
redner1998popular}%
\begin{APACrefauthors}%
Redner, S.%
\end{APACrefauthors}%
\unskip\
\newblock
\APACrefYearMonthDay{1998}{}{}.
\newblock
{\BBOQ}\APACrefatitle {How popular is your paper? An empirical study of the citation distribution} {How popular is your paper? an empirical study of the citation distribution}.{\BBCQ}
\newblock
\APACjournalVolNumPages{The European Physical Journal B---Condensed Matter and Complex Systems}{4}{2}{131--134}.
\PrintBackRefs{\CurrentBib}

\bibitem [\protect \citeauthoryear {%
Schubert%
\ \BBA {} Braun%
}{%
Schubert%
\ \BBA {} Braun%
}{%
{\protect \APACyear {1990}}%
}]{%
schubert1990international}
\APACinsertmetastar {%
schubert1990international}%
\begin{APACrefauthors}%
Schubert, A.%
\BCBT {}\ \BBA {} Braun, T.%
\end{APACrefauthors}%
\unskip\
\newblock
\APACrefYearMonthDay{1990}{}{}.
\newblock
{\BBOQ}\APACrefatitle {International collaboration in the sciences 1981--1985} {International collaboration in the sciences 1981--1985}.{\BBCQ}
\newblock
\APACjournalVolNumPages{Scientometrics}{19}{}{3--10}.
\PrintBackRefs{\CurrentBib}

\bibitem [\protect \citeauthoryear {%
Serrano%
, Bogun{\'a}%
\BCBL {}\ \BBA {} Vespignani%
}{%
Serrano%
\ \protect \BOthers {.}}{%
{\protect \APACyear {2009}}%
}]{%
serrano2009extracting}
\APACinsertmetastar {%
serrano2009extracting}%
\begin{APACrefauthors}%
Serrano, M\BPBI {\'A}.%
, Bogun{\'a}, M.%
\BCBL {}\ \BBA {} Vespignani, A.%
\end{APACrefauthors}%
\unskip\
\newblock
\APACrefYearMonthDay{2009}{}{}.
\newblock
{\BBOQ}\APACrefatitle {Extracting the multiscale backbone of complex weighted networks} {Extracting the multiscale backbone of complex weighted networks}.{\BBCQ}
\newblock
\APACjournalVolNumPages{Proceedings of the National Academy of Sciences}{106}{16}{6483--6488}.
\PrintBackRefs{\CurrentBib}

\bibitem [\protect \citeauthoryear {%
Sinatra%
, Wang%
, Deville%
, Song%
\BCBL {}\ \BBA {} Barab{\'a}si%
}{%
Sinatra%
\ \protect \BOthers {.}}{%
{\protect \APACyear {2016}}%
}]{%
sinatra2016quantifying}
\APACinsertmetastar {%
sinatra2016quantifying}%
\begin{APACrefauthors}%
Sinatra, R.%
, Wang, D.%
, Deville, P.%
, Song, C.%
\BCBL {}\ \BBA {} Barab{\'a}si, A\BHBI L.%
\end{APACrefauthors}%
\unskip\
\newblock
\APACrefYearMonthDay{2016}{}{}.
\newblock
{\BBOQ}\APACrefatitle {Quantifying the evolution of individual scientific impact} {Quantifying the evolution of individual scientific impact}.{\BBCQ}
\newblock
\APACjournalVolNumPages{Science}{354}{6312}{aaf5239}.
\PrintBackRefs{\CurrentBib}

\bibitem [\protect \citeauthoryear {%
Snijders%
}{%
Snijders%
}{%
{\protect \APACyear {2001}}%
}]{%
snijders2001statistical}
\APACinsertmetastar {%
snijders2001statistical}%
\begin{APACrefauthors}%
Snijders, T\BPBI A\BPBI B.%
\end{APACrefauthors}%
\unskip\
\newblock
\APACrefYearMonthDay{2001}{}{}.
\newblock
{\BBOQ}\APACrefatitle {The statistical evaluation of social network dynamics} {The statistical evaluation of social network dynamics}.{\BBCQ}
\newblock
\APACjournalVolNumPages{Sociological Methodology}{31}{1}{361--395}.
\PrintBackRefs{\CurrentBib}

\bibitem [\protect \citeauthoryear {%
Snijders%
, Ripley%
, B{\'o}da%
, V{\"o}r{\"o}s%
\BCBL {}\ \BBA {} Preciado%
}{%
Snijders%
\ \protect \BOthers {.}}{%
{\protect \APACyear {2025}}%
}]{%
snijders2025manual}
\APACinsertmetastar {%
snijders2025manual}%
\begin{APACrefauthors}%
Snijders, T\BPBI A\BPBI B.%
, Ripley, R\BPBI M.%
, B{\'o}da, Z.%
, V{\"o}r{\"o}s, A.%
\BCBL {}\ \BBA {} Preciado, P.%
\end{APACrefauthors}%
\unskip\
\newblock
\APACrefYearMonthDay{2025}{}{}.
\newblock
{\BBOQ}\APACrefatitle {Manual for {RSiena}} {Manual for {RSiena}}{\BBCQ}\ [\bibcomputersoftwaremanual].
\newblock
\APACaddressPublisher{Groningen, The Netherlands}{}.
\newblock
\begin{APACrefURL} \url{https://www.stats.ox.ac.uk/~snijders/siena/} \end{APACrefURL}
\newblock
\APACrefnote{R package version 1.5.0}
\PrintBackRefs{\CurrentBib}

\bibitem [\protect \citeauthoryear {%
Snijders%
, Steglich%
\BCBL {}\ \BBA {} Schweinberger%
}{%
Snijders%
\ \protect \BOthers {.}}{%
{\protect \APACyear {2017}}%
}]{%
snijders2017modeling}
\APACinsertmetastar {%
snijders2017modeling}%
\begin{APACrefauthors}%
Snijders, T\BPBI A\BPBI B.%
, Steglich, C.%
\BCBL {}\ \BBA {} Schweinberger, M.%
\end{APACrefauthors}%
\unskip\
\newblock
\APACrefYearMonthDay{2017}{}{}.
\newblock
{\BBOQ}\APACrefatitle {Modeling the coevolution of networks and behavior} {Modeling the coevolution of networks and behavior}.{\BBCQ}
\newblock
\BIn{} \APACrefbtitle {Longitudinal Models in the Behavioral and Related Sciences} {Longitudinal models in the behavioral and related sciences}\ (\BPGS\ 41--71).
\newblock
\APACaddressPublisher{}{Routledge}.
\PrintBackRefs{\CurrentBib}

\bibitem [\protect \citeauthoryear {%
Spence%
}{%
Spence%
}{%
{\protect \APACyear {1978}}%
}]{%
spence1978job}
\APACinsertmetastar {%
spence1978job}%
\begin{APACrefauthors}%
Spence, M.%
\end{APACrefauthors}%
\unskip\
\newblock
\APACrefYearMonthDay{1978}{}{}.
\newblock
{\BBOQ}\APACrefatitle {Job market signaling} {Job market signaling}.{\BBCQ}
\newblock
\BIn{} \APACrefbtitle {Uncertainty in Economics} {Uncertainty in economics}\ (\BPGS\ 281--306).
\newblock
\APACaddressPublisher{}{Elsevier}.
\PrintBackRefs{\CurrentBib}

\bibitem [\protect \citeauthoryear {%
Stadler%
}{%
Stadler%
}{%
{\protect \APACyear {2017}}%
}]{%
stadler2017country}
\APACinsertmetastar {%
stadler2017country}%
\begin{APACrefauthors}%
Stadler, K.%
\end{APACrefauthors}%
\unskip\
\newblock
\APACrefYearMonthDay{2017}{}{}.
\newblock
{\BBOQ}\APACrefatitle {The country converter coco -- a {Python} package for converting country names between different classification schemes} {The country converter coco -- a {Python} package for converting country names between different classification schemes}.{\BBCQ}
\newblock
\APACjournalVolNumPages{Journal of Open Source Software}{2}{16}{332}.
\PrintBackRefs{\CurrentBib}

\bibitem [\protect \citeauthoryear {%
Tenzer%
\ \BBA {} Pudelko%
}{%
Tenzer%
\ \BBA {} Pudelko%
}{%
{\protect \APACyear {2013}}%
}]{%
tenzer2013leading}
\APACinsertmetastar {%
tenzer2013leading}%
\begin{APACrefauthors}%
Tenzer, H.%
\BCBT {}\ \BBA {} Pudelko, M.%
\end{APACrefauthors}%
\unskip\
\newblock
\APACrefYearMonthDay{2013}{}{}.
\newblock
{\BBOQ}\APACrefatitle {Leading across language barriers: Strategies to mitigate negative language-induced emotions in {MNCs}} {Leading across language barriers: Strategies to mitigate negative language-induced emotions in {MNCs}}.{\BBCQ}
\newblock
\BIn{} \APACrefbtitle {Academy of Management Proceedings} {Academy of management proceedings}\ (\BVOL\ 2013, \BPG~12852).
\PrintBackRefs{\CurrentBib}

\bibitem [\protect \citeauthoryear {%
Tenzer%
, Pudelko%
\BCBL {}\ \BBA {} Harzing%
}{%
Tenzer%
\ \protect \BOthers {.}}{%
{\protect \APACyear {2014}}%
}]{%
tenzer2014impact}
\APACinsertmetastar {%
tenzer2014impact}%
\begin{APACrefauthors}%
Tenzer, H.%
, Pudelko, M.%
\BCBL {}\ \BBA {} Harzing, A\BHBI W.%
\end{APACrefauthors}%
\unskip\
\newblock
\APACrefYearMonthDay{2014}{}{}.
\newblock
{\BBOQ}\APACrefatitle {The impact of language barriers on trust formation in multinational teams} {The impact of language barriers on trust formation in multinational teams}.{\BBCQ}
\newblock
\APACjournalVolNumPages{Journal of International Business Studies}{45}{}{508--535}.
\PrintBackRefs{\CurrentBib}

\bibitem [\protect \citeauthoryear {%
Teorell%
\ \protect \BOthers {.}}{%
Teorell%
\ \protect \BOthers {.}}{%
{\protect \APACyear {2025}}%
}]{%
teorell2025qog}
\APACinsertmetastar {%
teorell2025qog}%
\begin{APACrefauthors}%
Teorell, J.%
, Sundstr{\"o}m, A.%
, Holmberg, S.%
, Rothstein, B.%
, Alvarado Pach{\'o}n, N.%
\BCBL {}\ \BBA {} others%
\end{APACrefauthors}%
\unskip\
\newblock
\APACrefYearMonthDay{2025}{}{}.
\newblock
\APACrefbtitle {The Quality of Government Standard Dataset, version {Jan25}.} {The Quality of Government Standard Dataset, version {Jan25}.}
\newblock
\APAChowpublished {University of Gothenburg: The Quality of Government Institute, \url{https://www.gu.se/en/quality-government}}.
\PrintBackRefs{\CurrentBib}

\bibitem [\protect \citeauthoryear {%
Thelwall%
\ \protect \BOthers {.}}{%
Thelwall%
\ \protect \BOthers {.}}{%
{\protect \APACyear {2023}}%
}]{%
thelwall2023coauthored}
\APACinsertmetastar {%
thelwall2023coauthored}%
\begin{APACrefauthors}%
Thelwall, M.%
, Kousha, K.%
, Abdoli, M.%
, Stuart, E.%
, Makita, M.%
, Wilson, P.%
\BCBL {}\ \BBA {} Levitt, J.%
\end{APACrefauthors}%
\unskip\
\newblock
\APACrefYearMonthDay{2023}{}{}.
\newblock
{\BBOQ}\APACrefatitle {Why are coauthored academic articles more cited: Higher quality or larger audience?} {Why are coauthored academic articles more cited: Higher quality or larger audience?}{\BBCQ}
\newblock
\APACjournalVolNumPages{Journal of the Association for Information Science and Technology}{74}{7}{791--810}.
\PrintBackRefs{\CurrentBib}

\bibitem [\protect \citeauthoryear {%
Tomkins%
, Zhang%
\BCBL {}\ \BBA {} Heavlin%
}{%
Tomkins%
\ \protect \BOthers {.}}{%
{\protect \APACyear {2017}}%
}]{%
tomkins2017reviewer}
\APACinsertmetastar {%
tomkins2017reviewer}%
\begin{APACrefauthors}%
Tomkins, A.%
, Zhang, M.%
\BCBL {}\ \BBA {} Heavlin, W\BPBI D.%
\end{APACrefauthors}%
\unskip\
\newblock
\APACrefYearMonthDay{2017}{}{}.
\newblock
{\BBOQ}\APACrefatitle {Reviewer bias in single- versus double-blind peer review} {Reviewer bias in single- versus double-blind peer review}.{\BBCQ}
\newblock
\APACjournalVolNumPages{Proceedings of the National Academy of Sciences}{114}{48}{12708--12713}.
\PrintBackRefs{\CurrentBib}

\bibitem [\protect \citeauthoryear {%
Uzzi%
, Mukherjee%
, Stringer%
\BCBL {}\ \BBA {} Jones%
}{%
Uzzi%
\ \protect \BOthers {.}}{%
{\protect \APACyear {2013}}%
}]{%
uzzi2013atypical}
\APACinsertmetastar {%
uzzi2013atypical}%
\begin{APACrefauthors}%
Uzzi, B.%
, Mukherjee, S.%
, Stringer, M.%
\BCBL {}\ \BBA {} Jones, B.%
\end{APACrefauthors}%
\unskip\
\newblock
\APACrefYearMonthDay{2013}{}{}.
\newblock
{\BBOQ}\APACrefatitle {Atypical combinations and scientific impact} {Atypical combinations and scientific impact}.{\BBCQ}
\newblock
\APACjournalVolNumPages{Science}{342}{6157}{468--472}.
\PrintBackRefs{\CurrentBib}

\bibitem [\protect \citeauthoryear {%
Vakilian%
, Yeop~Majlis%
\BCBL {}\ \BBA {} Mousavi%
}{%
Vakilian%
\ \protect \BOthers {.}}{%
{\protect \APACyear {2015}}%
}]{%
vakilian2015bibliometric}
\APACinsertmetastar {%
vakilian2015bibliometric}%
\begin{APACrefauthors}%
Vakilian, M.%
, Yeop~Majlis, B.%
\BCBL {}\ \BBA {} Mousavi, M.%
\end{APACrefauthors}%
\unskip\
\newblock
\APACrefYearMonthDay{2015}{}{}.
\newblock
{\BBOQ}\APACrefatitle {A bibliometric analysis of lab-on-a-chip research from 2001 to 2013} {A bibliometric analysis of lab-on-a-chip research from 2001 to 2013}.{\BBCQ}
\newblock
\APACjournalVolNumPages{Scientometrics}{105}{}{789--804}.
\PrintBackRefs{\CurrentBib}

\bibitem [\protect \citeauthoryear {%
Wagner%
}{%
Wagner%
}{%
{\protect \APACyear {2024}}%
}]{%
wagner2024science}
\APACinsertmetastar {%
wagner2024science}%
\begin{APACrefauthors}%
Wagner, C\BPBI S.%
\end{APACrefauthors}%
\unskip\
\newblock
\APACrefYearMonthDay{2024}{}{}.
\newblock
{\BBOQ}\APACrefatitle {Science and the nation-state: What {China}'s experience reveals about the role of policy in science} {Science and the nation-state: What {China}'s experience reveals about the role of policy in science}.{\BBCQ}
\newblock
\APACjournalVolNumPages{Science and Public Policy}{}{}{scae034}.
\PrintBackRefs{\CurrentBib}

\bibitem [\protect \citeauthoryear {%
Wagner%
\ \BBA {} Leydesdorff%
}{%
Wagner%
\ \BBA {} Leydesdorff%
}{%
{\protect \APACyear {2005}}%
}]{%
wagner2005network}
\APACinsertmetastar {%
wagner2005network}%
\begin{APACrefauthors}%
Wagner, C\BPBI S.%
\BCBT {}\ \BBA {} Leydesdorff, L.%
\end{APACrefauthors}%
\unskip\
\newblock
\APACrefYearMonthDay{2005}{}{}.
\newblock
{\BBOQ}\APACrefatitle {Network structure, self-organization, and the growth of international collaboration in science} {Network structure, self-organization, and the growth of international collaboration in science}.{\BBCQ}
\newblock
\APACjournalVolNumPages{Research Policy}{34}{10}{1608--1618}.
\PrintBackRefs{\CurrentBib}

\bibitem [\protect \citeauthoryear {%
Wagner%
, Park%
\BCBL {}\ \BBA {} Leydesdorff%
}{%
Wagner%
\ \protect \BOthers {.}}{%
{\protect \APACyear {2015}}%
}]{%
wagner2015continuing}
\APACinsertmetastar {%
wagner2015continuing}%
\begin{APACrefauthors}%
Wagner, C\BPBI S.%
, Park, H\BPBI W.%
\BCBL {}\ \BBA {} Leydesdorff, L.%
\end{APACrefauthors}%
\unskip\
\newblock
\APACrefYearMonthDay{2015}{}{}.
\newblock
{\BBOQ}\APACrefatitle {The continuing growth of global cooperation networks in research: A conundrum for national governments} {The continuing growth of global cooperation networks in research: A conundrum for national governments}.{\BBCQ}
\newblock
\APACjournalVolNumPages{PLOS ONE}{10}{7}{e0131816}.
\PrintBackRefs{\CurrentBib}

\bibitem [\protect \citeauthoryear {%
Wagner%
, Whetsell%
, Baas%
\BCBL {}\ \BBA {} Jonkers%
}{%
Wagner%
\ \protect \BOthers {.}}{%
{\protect \APACyear {2018}}%
}]{%
wagner2018openness}
\APACinsertmetastar {%
wagner2018openness}%
\begin{APACrefauthors}%
Wagner, C\BPBI S.%
, Whetsell, T.%
, Baas, J.%
\BCBL {}\ \BBA {} Jonkers, K.%
\end{APACrefauthors}%
\unskip\
\newblock
\APACrefYearMonthDay{2018}{}{}.
\newblock
{\BBOQ}\APACrefatitle {Openness and impact of leading scientific countries} {Openness and impact of leading scientific countries}.{\BBCQ}
\newblock
\APACjournalVolNumPages{Frontiers in Research Metrics and Analytics}{3}{}{10}.
\PrintBackRefs{\CurrentBib}

\bibitem [\protect \citeauthoryear {%
Wagner%
, Whetsell%
\BCBL {}\ \BBA {} Leydesdorff%
}{%
Wagner%
\ \protect \BOthers {.}}{%
{\protect \APACyear {2017}}%
}]{%
wagner2017growth}
\APACinsertmetastar {%
wagner2017growth}%
\begin{APACrefauthors}%
Wagner, C\BPBI S.%
, Whetsell, T\BPBI A.%
\BCBL {}\ \BBA {} Leydesdorff, L.%
\end{APACrefauthors}%
\unskip\
\newblock
\APACrefYearMonthDay{2017}{}{}.
\newblock
{\BBOQ}\APACrefatitle {Growth of international collaboration in science: revisiting six specialties} {Growth of international collaboration in science: revisiting six specialties}.{\BBCQ}
\newblock
\APACjournalVolNumPages{Scientometrics}{110}{}{1633--1652}.
\PrintBackRefs{\CurrentBib}

\bibitem [\protect \citeauthoryear {%
Wagner%
, Whetsell%
\BCBL {}\ \BBA {} Mukherjee%
}{%
Wagner%
\ \protect \BOthers {.}}{%
{\protect \APACyear {2019}}%
}]{%
wagner2019international}
\APACinsertmetastar {%
wagner2019international}%
\begin{APACrefauthors}%
Wagner, C\BPBI S.%
, Whetsell, T\BPBI A.%
\BCBL {}\ \BBA {} Mukherjee, S.%
\end{APACrefauthors}%
\unskip\
\newblock
\APACrefYearMonthDay{2019}{}{}.
\newblock
{\BBOQ}\APACrefatitle {International research collaboration: Novelty, conventionality, and atypicality in knowledge recombination} {International research collaboration: Novelty, conventionality, and atypicality in knowledge recombination}.{\BBCQ}
\newblock
\APACjournalVolNumPages{Research Policy}{48}{5}{1260--1270}.
\PrintBackRefs{\CurrentBib}

\bibitem [\protect \citeauthoryear {%
Waltman%
}{%
Waltman%
}{%
{\protect \APACyear {2016}}%
}]{%
waltman2016review}
\APACinsertmetastar {%
waltman2016review}%
\begin{APACrefauthors}%
Waltman, L.%
\end{APACrefauthors}%
\unskip\
\newblock
\APACrefYearMonthDay{2016}{}{}.
\newblock
{\BBOQ}\APACrefatitle {A review of the literature on citation impact indicators} {A review of the literature on citation impact indicators}.{\BBCQ}
\newblock
\APACjournalVolNumPages{Journal of Informetrics}{10}{2}{365--391}.
\PrintBackRefs{\CurrentBib}

\bibitem [\protect \citeauthoryear {%
Waltman%
, Tijssen%
\BCBL {}\ \BBA {} van~Eck%
}{%
Waltman%
\ \protect \BOthers {.}%
}{%
{\protect \APACyear {2011}}%
}]{%
waltman2011globalisation}
\APACinsertmetastar {%
waltman2011globalisation}%
\begin{APACrefauthors}%
Waltman, L.%
, Tijssen, R\BPBI J\BPBI W.%
\BCBL {}\ \BBA {} van~Eck, N\BPBI J.%
\end{APACrefauthors}%
\unskip\
\newblock
\APACrefYearMonthDay{2011}{}{}.
\newblock
{\BBOQ}\APACrefatitle {Globalisation of science in kilometres} {Globalisation of science in kilometres}.{\BBCQ}
\newblock
\APACjournalVolNumPages{Journal of Informetrics}{5}{4}{574--582}.
\PrintBackRefs{\CurrentBib}

\bibitem [\protect \citeauthoryear {%
Waltman%
\ \BBA {} van~Eck%
}{%
Waltman%
\ \BBA {} van~Eck%
}{%
{\protect \APACyear {2015}}%
}]{%
waltman2015field}
\APACinsertmetastar {%
waltman2015field}%
\begin{APACrefauthors}%
Waltman, L.%
\BCBT {}\ \BBA {} van~Eck, N\BPBI J.%
\end{APACrefauthors}%
\unskip\
\newblock
\APACrefYearMonthDay{2015}{}{}.
\newblock
{\BBOQ}\APACrefatitle {Field-normalized citation impact indicators and the choice of an appropriate counting method} {Field-normalized citation impact indicators and the choice of an appropriate counting method}.{\BBCQ}
\newblock
\APACjournalVolNumPages{Journal of Informetrics}{9}{4}{872--894}.
\PrintBackRefs{\CurrentBib}

\bibitem [\protect \citeauthoryear {%
Wang%
, Thijs%
\BCBL {}\ \BBA {} Gl{\"a}nzel%
}{%
Wang%
\ \protect \BOthers {.}}{%
{\protect \APACyear {2015}}%
}]{%
wang2015characteristics}
\APACinsertmetastar {%
wang2015characteristics}%
\begin{APACrefauthors}%
Wang, L.%
, Thijs, B.%
\BCBL {}\ \BBA {} Gl{\"a}nzel, W.%
\end{APACrefauthors}%
\unskip\
\newblock
\APACrefYearMonthDay{2015}{}{}.
\newblock
{\BBOQ}\APACrefatitle {Characteristics of international collaboration in sport sciences publications and its influence on citation impact} {Characteristics of international collaboration in sport sciences publications and its influence on citation impact}.{\BBCQ}
\newblock
\APACjournalVolNumPages{Scientometrics}{105}{}{843--862}.
\PrintBackRefs{\CurrentBib}

\bibitem [\protect \citeauthoryear {%
Wang%
\ \protect \BOthers {.}}{%
Wang%
\ \protect \BOthers {.}}{%
{\protect \APACyear {2023}}%
}]{%
wang2023effect}
\APACinsertmetastar {%
wang2023effect}%
\begin{APACrefauthors}%
Wang, Y.%
, Li, N.%
, Zhang, B.%
, Huang, Q.%
, Wu, J.%
\BCBL {}\ \BBA {} Wang, Y.%
\end{APACrefauthors}%
\unskip\
\newblock
\APACrefYearMonthDay{2023}{}{}.
\newblock
{\BBOQ}\APACrefatitle {The effect of structural holes on producing novel and disruptive research in physics} {The effect of structural holes on producing novel and disruptive research in physics}.{\BBCQ}
\newblock
\APACjournalVolNumPages{Scientometrics}{128}{3}{1801--1823}.
\PrintBackRefs{\CurrentBib}

\bibitem [\protect \citeauthoryear {%
Whetsell%
}{%
Whetsell%
}{%
{\protect \APACyear {2023}}%
}]{%
whetsell2023democratic}
\APACinsertmetastar {%
whetsell2023democratic}%
\begin{APACrefauthors}%
Whetsell, T\BPBI A.%
\end{APACrefauthors}%
\unskip\
\newblock
\APACrefYearMonthDay{2023}{}{}.
\newblock
{\BBOQ}\APACrefatitle {Democratic governance and global science: A longitudinal analysis of the international research collaboration network} {Democratic governance and global science: A longitudinal analysis of the international research collaboration network}.{\BBCQ}
\newblock
\APACjournalVolNumPages{PLOS ONE}{18}{6}{e0287058}.
\PrintBackRefs{\CurrentBib}

\bibitem [\protect \citeauthoryear {%
Whetsell%
, Dimand%
, Jonkers%
, Baas%
\BCBL {}\ \BBA {} Wagner%
}{%
Whetsell%
\ \protect \BOthers {.}}{%
{\protect \APACyear {2021}}%
}]{%
whetsell2021democracy}
\APACinsertmetastar {%
whetsell2021democracy}%
\begin{APACrefauthors}%
Whetsell, T\BPBI A.%
, Dimand, A\BHBI M.%
, Jonkers, K.%
, Baas, J.%
\BCBL {}\ \BBA {} Wagner, C\BPBI S.%
\end{APACrefauthors}%
\unskip\
\newblock
\APACrefYearMonthDay{2021}{}{}.
\newblock
{\BBOQ}\APACrefatitle {Democracy, complexity, and science: Exploring structural sources of national scientific performance} {Democracy, complexity, and science: Exploring structural sources of national scientific performance}.{\BBCQ}
\newblock
\APACjournalVolNumPages{Science and Public Policy}{48}{5}{697--711}.
\PrintBackRefs{\CurrentBib}

\bibitem [\protect \citeauthoryear {%
Whetsell%
\ \BBA {} Sidorova%
}{%
Whetsell%
\ \BBA {} Sidorova%
}{%
{\protect \APACyear {2024}}%
}]{%
whetsell2024academic}
\APACinsertmetastar {%
whetsell2024academic}%
\begin{APACrefauthors}%
Whetsell, T\BPBI A.%
\BCBT {}\ \BBA {} Sidorova, J.%
\end{APACrefauthors}%
\unskip\
\newblock
\APACrefYearMonthDay{2024}{}{}.
\newblock
{\BBOQ}\APACrefatitle {Academic Freedom and International Research Collaboration: A Longitudinal Analysis of Global Network Evolution} {Academic freedom and international research collaboration: A longitudinal analysis of global network evolution}.{\BBCQ}
\newblock
\APACjournalVolNumPages{arXiv preprint arXiv:2407.03968}{}{}{}.
\PrintBackRefs{\CurrentBib}

\bibitem [\protect \citeauthoryear {%
Wray%
}{%
Wray%
}{%
{\protect \APACyear {2002}}%
}]{%
Wray_2002}
\APACinsertmetastar {%
Wray_2002}%
\begin{APACrefauthors}%
Wray, K\BPBI B.%
\end{APACrefauthors}%
\unskip\
\newblock
\APACrefYearMonthDay{2002}{}{}.
\newblock
{\BBOQ}\APACrefatitle {The Epistemic Significance of Collaborative Research} {The epistemic significance of collaborative research}.{\BBCQ}
\newblock
\APACjournalVolNumPages{Philosophy of Science}{69}{1}{150--168}.
\newblock
\begin{APACrefDOI} \doi{10.1086/338946} \end{APACrefDOI}
\PrintBackRefs{\CurrentBib}

\bibitem [\protect \citeauthoryear {%
Wuchty%
, Jones%
\BCBL {}\ \BBA {} Uzzi%
}{%
Wuchty%
\ \protect \BOthers {.}}{%
{\protect \APACyear {2007}}%
}]{%
wuchty2007increasing}
\APACinsertmetastar {%
wuchty2007increasing}%
\begin{APACrefauthors}%
Wuchty, S.%
, Jones, B\BPBI F.%
\BCBL {}\ \BBA {} Uzzi, B.%
\end{APACrefauthors}%
\unskip\
\newblock
\APACrefYearMonthDay{2007}{}{}.
\newblock
{\BBOQ}\APACrefatitle {The increasing dominance of teams in production of knowledge} {The increasing dominance of teams in production of knowledge}.{\BBCQ}
\newblock
\APACjournalVolNumPages{Science}{316}{5827}{1036--1039}.
\PrintBackRefs{\CurrentBib}

\bibitem [\protect \citeauthoryear {%
Wuestman%
, Hoekman%
\BCBL {}\ \BBA {} Frenken%
}{%
Wuestman%
\ \protect \BOthers {.}}{%
{\protect \APACyear {2019}}%
}]{%
wuestman2019geography}
\APACinsertmetastar {%
wuestman2019geography}%
\begin{APACrefauthors}%
Wuestman, M\BPBI L.%
, Hoekman, J.%
\BCBL {}\ \BBA {} Frenken, K.%
\end{APACrefauthors}%
\unskip\
\newblock
\APACrefYearMonthDay{2019}{}{}.
\newblock
{\BBOQ}\APACrefatitle {The geography of scientific citations} {The geography of scientific citations}.{\BBCQ}
\newblock
\APACjournalVolNumPages{Research Policy}{48}{7}{1771--1780}.
\PrintBackRefs{\CurrentBib}

\bibitem [\protect \citeauthoryear {%
Yadav%
, McHale%
\BCBL {}\ \BBA {} O'Neill%
}{%
Yadav%
\ \protect \BOthers {.}}{%
{\protect \APACyear {2023}}%
}]{%
yadav2023does}
\APACinsertmetastar {%
yadav2023does}%
\begin{APACrefauthors}%
Yadav, A.%
, McHale, J.%
\BCBL {}\ \BBA {} O'Neill, S.%
\end{APACrefauthors}%
\unskip\
\newblock
\APACrefYearMonthDay{2023}{}{}.
\newblock
{\BBOQ}\APACrefatitle {How does co-authoring with a star affect scientists' productivity? Evidence from small open economies} {How does co-authoring with a star affect scientists' productivity? evidence from small open economies}.{\BBCQ}
\newblock
\APACjournalVolNumPages{Research Policy}{52}{1}{104660}.
\PrintBackRefs{\CurrentBib}

\bibitem [\protect \citeauthoryear {%
Yan%
\ \BBA {} Sugimoto%
}{%
Yan%
\ \BBA {} Sugimoto%
}{%
{\protect \APACyear {2011}}%
}]{%
yan2011institutional}
\APACinsertmetastar {%
yan2011institutional}%
\begin{APACrefauthors}%
Yan, E.%
\BCBT {}\ \BBA {} Sugimoto, C\BPBI R.%
\end{APACrefauthors}%
\unskip\
\newblock
\APACrefYearMonthDay{2011}{}{}.
\newblock
{\BBOQ}\APACrefatitle {Institutional interactions: Exploring social, cognitive, and geographic relationships between institutions as demonstrated through citation networks} {Institutional interactions: Exploring social, cognitive, and geographic relationships between institutions as demonstrated through citation networks}.{\BBCQ}
\newblock
\APACjournalVolNumPages{Journal of the American Society for Information Science and Technology}{62}{8}{1498--1514}.
\PrintBackRefs{\CurrentBib}

\bibitem [\protect \citeauthoryear {%
Zaheer%
\ \BBA {} Bell%
}{%
Zaheer%
\ \BBA {} Bell%
}{%
{\protect \APACyear {2005}}%
}]{%
zaheer2005benefiting}
\APACinsertmetastar {%
zaheer2005benefiting}%
\begin{APACrefauthors}%
Zaheer, A.%
\BCBT {}\ \BBA {} Bell, G\BPBI G.%
\end{APACrefauthors}%
\unskip\
\newblock
\APACrefYearMonthDay{2005}{}{}.
\newblock
{\BBOQ}\APACrefatitle {Benefiting from network position: firm capabilities, structural holes, and performance} {Benefiting from network position: firm capabilities, structural holes, and performance}.{\BBCQ}
\newblock
\APACjournalVolNumPages{Strategic Management Journal}{26}{9}{809--825}.
\PrintBackRefs{\CurrentBib}

\bibitem [\protect \citeauthoryear {%
Zhang%
, Ding%
, Wang%
\BCBL {}\ \BBA {} Liu%
}{%
Zhang%
\ \protect \BOthers {.}}{%
{\protect \APACyear {2022}}%
}]{%
zhang2022exploring}
\APACinsertmetastar {%
zhang2022exploring}%
\begin{APACrefauthors}%
Zhang, D.%
, Ding, W.%
, Wang, Y.%
\BCBL {}\ \BBA {} Liu, S.%
\end{APACrefauthors}%
\unskip\
\newblock
\APACrefYearMonthDay{2022}{}{}.
\newblock
{\BBOQ}\APACrefatitle {Exploring the role of international research collaboration in building {China}'s world-class universities} {Exploring the role of international research collaboration in building {China}'s world-class universities}.{\BBCQ}
\newblock
\APACjournalVolNumPages{Sustainability}{14}{6}{3487}.
\PrintBackRefs{\CurrentBib}

\end{thebibliography}
\end{document}